\documentclass[aps,prx,twocolumn,final,letterpaper]{revtex4}

\usepackage{appendix}
\usepackage{graphicx}   
\usepackage{import}                         
\usepackage{epstopdf}
\usepackage{amsmath} 
\usepackage{bm}
\usepackage{amssymb}
\usepackage{quotes}
\usepackage{transparent}
\usepackage{dcolumn}
\usepackage{multirow}
\usepackage{cancel} 
\usepackage{mdframed}
\usepackage{color}
\usepackage{bm}
\usepackage{dsfont}
\usepackage{slashed}
\usepackage{enumitem}
\usepackage{braket}

\begin{document}
\rmfamily

\title{Driven-dissipative phases and dynamics in non-Markovian nonlinear photonics}



\newcommand{\MITrle}{Research Laboratory of Electronics, Massachusetts Institute of Technology, Cambridge, MA 02139, USA}
\newcommand{\harvard}{Department of Physics, Harvard University, Cambridge, MA 02138, USA.}
\newcommand{\MITphysics}{Department of Physics, Massachusetts Institute of Technology, Cambridge, MA 02139, USA}




\author{Jamison Sloan,$^{1,\dagger}$ Nicholas Rivera,$^{2,3,\dagger}$ Marin Solja\v{c}i\'{c}$^{1,3}$}
\affiliation{\MITrle \\ \harvard \\ \MITphysics \\ $^\dagger$Denotes equal contribution.}

\begin{abstract} 

Interactions between photons (nonlinearities) enable a powerful form of control over the state of light. This control has enabled technologies such as light sources at new wavelengths, ultra-short optical pulses, frequency-comb metrology systems, even quantum light sources. 
Common to a wide variety of nonlinear optical technologies is an equilibrium between an energy source, such as an external laser, and dissipation, such as radiation loss or absorption. In the vast majority of these systems, the coupling between the system and the outside world (which leads to loss) is well-described as ``Markovian,'' meaning that the outside world has no memory of its past state. In this work, we introduce a class of driven-dissipative systems in which a nonlinear cavity experiences non-Markovian coupling to the outside world. In the classical regime, we show that these non-Markovian cavities can have extremely low thresholds for nonlinear effects, as well as self-pulsing instabilities at THz rates, and rich phase diagrams with alternating regions of stability and instability. In the quantum regime, we show how these system, when implemented on state-of-the-art platforms, can enable generation of strongly squeezed cavity states with intensity fluctuations that can be more than 15 dB below the classical limit, in contrast to the Markovian driven-dissipative cavity, in which the limit is 3 dB. In the regime of few-photon nonlinearity, such non-Markovian cavities can enable a deterministic protocol to generate Fock states of high order, which are long-desired, but still elusive at optical frequencies. We expect that exploiting non-Markovian couplings in nonlinear optics should in the future lead to even richer possibilities than those discussed here for both classical and quantum light manipulations. 

\end{abstract}

\maketitle

\section{Introduction}


Nonlinear systems are ubiquitous across scientific disciplines, exhibiting universal phenomena such as phase transitions, synchronization, pattern formation, and chaotic behavior \cite{kaplan1997understanding, nayfeh2008applied, thompson1990nonlinear}. Nonlinearity also plays a central role in optics, where materials with a nonlinear polarization response enable frequency conversion, field sensing, and ultrashort pulse generation \cite{boyd2020nonlinear, shen1984principles}. The invention of the laser quickly enabled the observation of many classical nonlinear effects, including harmonic generation \cite{franken1961generation}, soliton formation \cite{kivshar2003optical}, self-focusing \cite{kelley1965self}, self-phase modulation \cite{stolen1978self}, and optical parametric amplification, all of which are still intensely researched to this day.

Of particular importance in optics are so-called ``driven-dissipative'' systems. Such systems typically consist of a nonlinear optical resonator (or multiple resonators) driven by an external light source. The simultaneous presence of nonlinearity, dissipation, and external drive lead to striking classical effects such as cavity bistability \cite{drummond1980quantum, lugiato1984ii, gibbs1979optical, gibbs2012optical}, dissipative Kerr solitons in waveguides \cite{kippenberg2018dissipative, guo2017universal, kues2019quantum, guidry2022quantum, chembo2016quantum}, and optical parametric oscillation \cite{harris1969tunable}. Nonlinear optical systems also enable transformations of the quantum state of light \cite{drummond2014quantum}, enabling key applications in metrology and quantum information processing. Such systems have also been proposed as a platform to study collective behavior and phase transitions \cite{maghrebi2016nonequilibrium, foss2017emergent, zhang2021driven, le2013steady}.




In the vast majority of driven-dissipative systems, the coupling between the system and its environment is assumed to be independent of frequency over the bandwidths of interest. This equivalently means that the system's interaction with its environment is assumed to be instantaneous, or ``Markovian.'' In such systems, the outside world (environment) retains no memory of its prior state, meaning that the outside world interaction is in some sense uncorrelated. Given the importance of correlations in interacting systems for realizing useful behaviors, it is surprising that the regime of strong nonlinearity and non-Markovian dissipation remains largely unexplored. This is especially relevant given that a wide variety of platforms (free space optical filters, fiber systems, photonic crystals, integrated nanophotonics, etc.) can be readily engineered to provide exactly the types of strong frequency-dependent couplings which undermine the Markovian assumption. Thus, while the Markovian assumption is well respected in many systems, it also places considerable limitations on the space of possible driven-dissipative architectures, and the corresponding functionalities which can be realized. 


In this work, we develop the physics of driven-dissipative systems with non-Markovian couplings to the environment. We introduce a general class of models which consist of an intensity-dependent (Kerr) nonlinear optical resonance coupled to a number of non-Markovian continuum channels, and solve for their classical and quantum state dynamics. We show that the resonance-frequency-dependent loss in these systems enables new driven-dissipative phases. In the classical domain, we identify low-threshold bistability, as well as self-pulsing instabilities that enable passive modulation of incoming light. In the quantum domain, the combination of bistability and non-Markovian loss enables the natural generation of strongly intensity-squeezed cavity states which are maintained in the steady-state by an external drive. We show how this effect arises from the physics of ``sharp loss'' which is unique to non-Markovian dissipation, and that mechanism can produce cavity states with intensity fluctuations more than 10 dB below the shot noise limit. In systems with particularly strong nonlinearities, this behavior enables the generation of high order Fock states, which have remained unrealized at optical frequencies, despite their importance for metrology and quantum information.
This introduction of non-Markovian coupling into the already rich space of driven-dissipative systems constitutes a new degree of control which can be used to engineer the quantum correlations of light by modifying a seemingly classical element such as frequency-dependent coupling.

\section{Theory}

\begin{figure}
    \centering
    \includegraphics{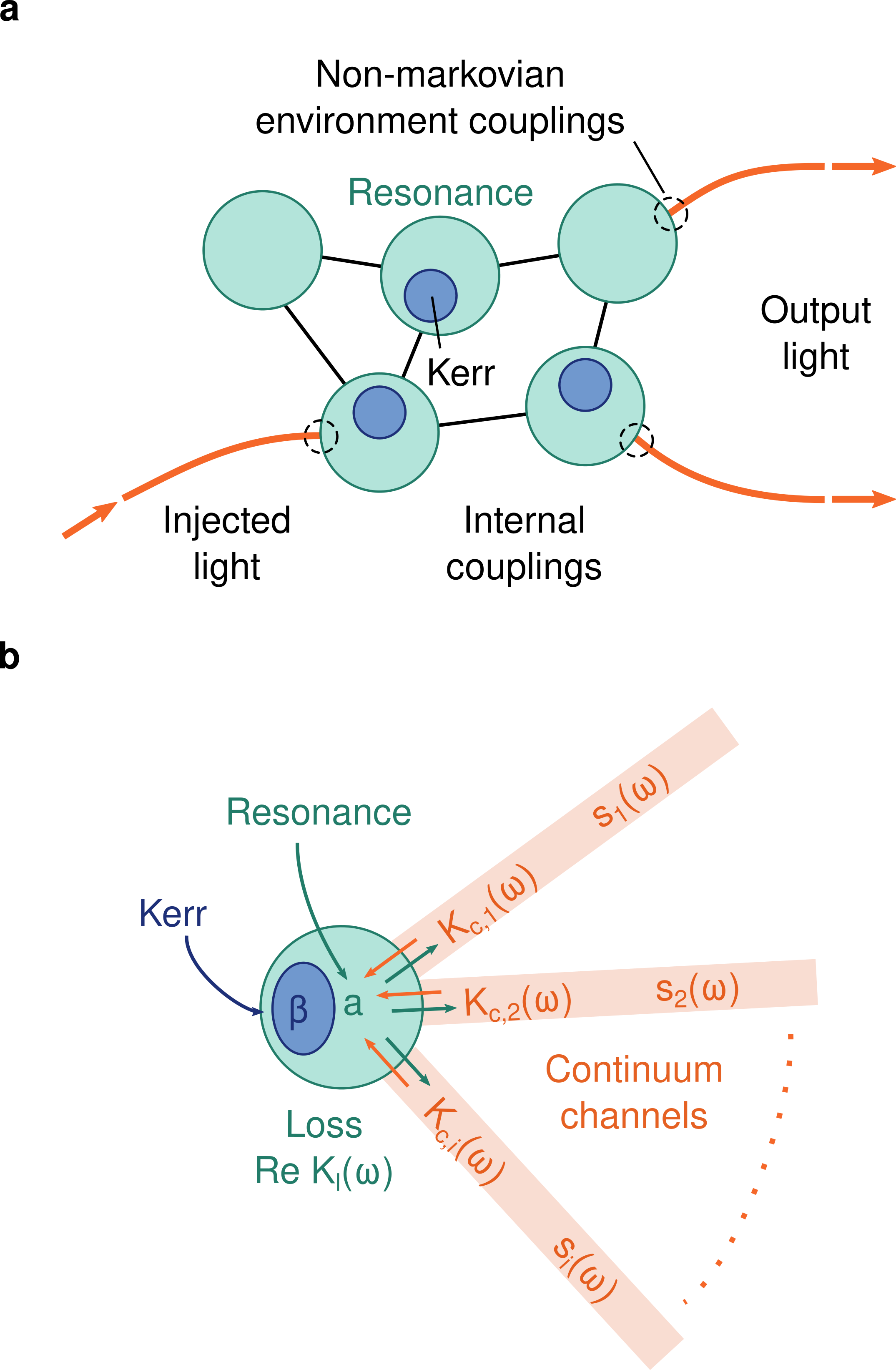}
    \caption{\textbf{General framework for non-Markovian driven-dissipative systems.} (a) General driven-dissipative system with non-Markovian couplings to the environment, and internal resonances which may contain Kerr nonlinearity. (b) Model primarily considered in this work (Eq.~\ref{eq:general_eom}), consisting of a nonlinear resonance $a$ coupled to one or more continuum channels $i$ through coupling functions $K_{c,i}(\omega)$. These reservoir couplings give the resonance a dispersive loss $K_l(\omega)$.}
    \label{fig:schematic}
\end{figure}

Our results are based on a general quantum optical theory of nonlinear resonances with frequency-dependent (non-Markovian) couplings to the environment, enabling the description of a highly general class of driven-dissipative systems (Fig.~\ref{fig:schematic}a). In this work, we focus more specifically on the class of systems comprising two key elements:  (1) frequency-dependent enivornment coupling which creates a resonance-frequency-dependent loss, and (2) Kerr nonlinearity which causes the frequency of resonance $a$ to depend on the number of photons $n$ in the resonance (Fig.~\ref{fig:schematic}b). 



We first describe the quantum optical theory of dispersive dissipation in linear resonators. 
To do so, we consider a resonance $a$ coupled to continuum channels (reservoirs) labeled $i$ (Fig. \ref{fig:schematic}b). 
We assume the resonance $a$ exchanges excitations with the reservoir fields $s_i$ via frequency-dependent coupling functions $K_{c,i}(\omega)$. 
This action of the coupling on the cavity is encoded in a frequency-dependent loss function $K_l(\omega)$: its real part gives the loss rate, and its imaginary part gives the frequency shift. Importantly, $K_{c,i}$ and $K_l$ are not independent, but rather constrained by a Kramers-Kronig relation, 
from which it follows that $2\,\text{Re}\,K_l(\omega) = \sum_i |K_{c,i}(\omega)|^2$. 



Kerr nonlinearity in the resonator equips the mode $a$ with an intensity-dependent resonance frequency $\omega(n) = \omega_a + \beta n$. Here, $\omega_a$ is the bare resonance frequency of $a$, $n$ is the cavity photon number, and $\beta$ is the Kerr frequency shift which results from adding one photon to the cavity. When the dispersive dissipation is simultaneously present, the resonance $a$ obeys the following Heisenberg-Langevin equation of motion:
\begin{equation}
\begin{split}
    \dot{a} = -i\omega_a a - \underbrace{i\beta(a^\dagger a) a}_{\text{Kerr}} & -\underbrace{\int dt'\,K_l(t-t')a(t')}_{\text{Cavity field damping with memory}} \\
        &+ \underbrace{\sum_i \int dt'\, K_{c,i}(t-t')s_i(t')}_{\text{Coupling of input fields with memory}}.
\end{split}
    \label{eq:general_eom}
\end{equation}
Here, $K_{l,c}(t - t')$ are the time-domain loss (coupling) kernels, related to the frequency-domain functions by $K_{l,c}(\tau) = \frac{1}{2\pi}\int d\omega\, e^{-i\omega\tau}K_{l,c}(\omega)$. 

This dissipation term is balanced by the presence of a quantum operator-valued input term. The input fields are normalized such that $[s_i(t), s_j^\dagger(t')] = \delta_{ij}\delta(t - t')$. Additionally, they can be decomposed into a sum of mean-field (c-number) and quantum fluctuation (operator) contributions $s_i = \braket{s_i} + \delta s_i$. The c-numbers give forcing terms which drive the mean-field dynamics of $a$, while the operator valued fluctuations generate a Langevin force term $F(t) \equiv \sum_i \int dt'\, K_{c,i}(t - t') \delta s_i(t')$ which adds non-Markovian fluctuations into the cavity.
It follows from the above that for vacuum reservoirs, the frequency-domain correlations of $F$ are $\braket{F(\omega)F^\dagger(\omega')} = 2\pi \cdot 2\,\text{Re} K_l(\omega) \delta(\omega - \omega')$. Thus, the correlations are local in frequency space (i.e., fluctuations at different frequencies are not correlated with one another), but with a magnitude that depends on frequency through the loss rate. It can be shown (see S.I.) that the stated correlation functions of the Langevin force lead to the preservation of the equal-time commutation relation of the cavity field, namely $[a(t), a^\dagger(t)] = 1$, indicating the self-consistency of the theory. Moreover, it can be shown that when the bandwidth of $K_c(\omega)$ is large, the dynamics revert back to the standard case with frequency-independent couplings.

\begin{figure*}
    \centering
    \includegraphics{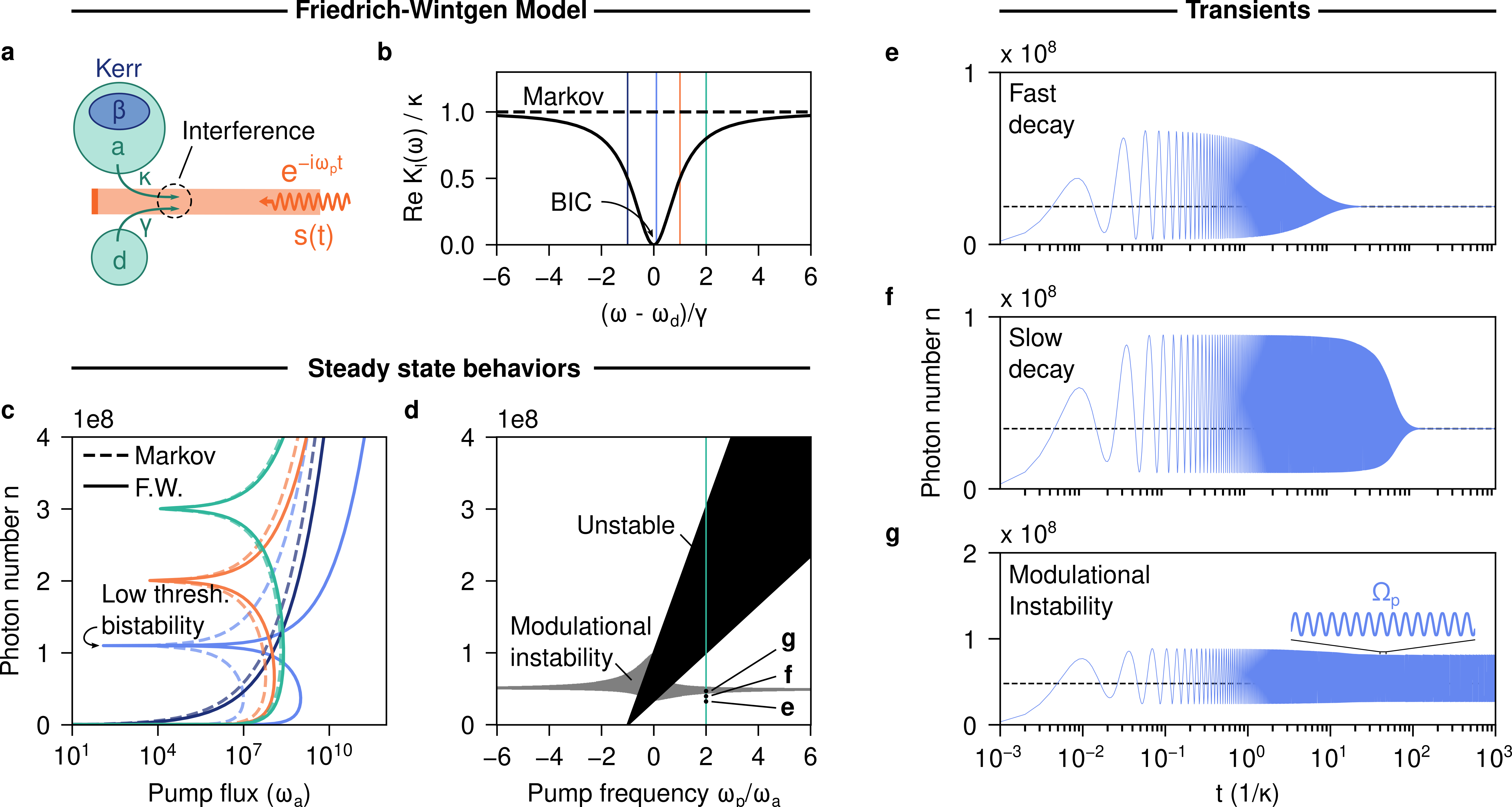}
    \caption{\textbf{mean-field behavior of a non-Markovian nonlinear driven-dissipative cavity.} (a) Schematic of a system described by a nonlinear Friedrich-Wintgen (F.W.) model which consists of two resonances $a$ and $d$ coupled to a common continuum $s$. (b) Dispersive loss profile $\text{Re}\,K_l(\omega)$ of the nonlinear Friedrich-Wintgen (F.W.) model. Destructive interference in the loss channels causes the dip in the loss experienced by $a$. Parameters used are $\kappa/\omega_a = 10^{-4}$, $\gamma/\omega_a = 10^{-2}$, and $\omega_d = \omega_a + \gamma$. (c) steady-state cavity photon number in the presence of a monochromatic pump at several different frequencies $\omega_p$ (marked as vertical lines in (b)). Dashed lines indicate behavior in the Markovian model, while solid lines indicate behavior in the non-Markovian model. Both models exhibit bistability for some detunings. The level of nonlinearity is set to $\beta/\omega_a = 10^{-10}$. (d) Phase diagram of the nonlinear F.W. model. Black region indicates conventional cavity bistability, while grey region indicates self-pulsing due to modulational instability (MI). 
    (e-g) Transient behaviors of the system operating at different pump rates corresponding to different steady-state photon numbers $n$ shown by horizontal dashed lines, and also marked in (d). Decay to a steady-state via relaxation oscillations is seen in (e, f), while self-pulsing instability is seen in (g).}
    \label{fig:mean_field}
\end{figure*}

This quantum framework of dispersive dissipation is relevant to a diverse array of physical systems. 
In fact, for any system where the density of states of the outside world (the loss channels) is significantly frequency-dependent this formalism can be applied. As examples, there are many photonic structures which can be used as sharp frequency-dependent elements, such as Bragg gratings, filters with Lorentzian and Fano profiles, photonic crystal mirrors, and so on. Additionally, schemes involving delay lines, or time-multiplexing \cite{leefmans2021topological}, could also be used to realize non-Markovian dissipation. The key point of our work is that nonlinearity can "interact" with these resonance-frequency-dependent losses to lead to new phenomena, especially in the presence of driving.



\section{Results}

As a minimal example which exhibits many important features, we introduce the nonlinear Friedrich-Wintgen (F.W.) model. The F.W. model is a temporal coupled mode theory (TCMT) model which describes two resonances $a$ and $d$ coupled to a common continuum (Fig.~\ref{fig:mean_field}a) \cite{friedrich1985interfering}. The losses of the eigenmodes of this model depend strongly on the relative frequencies of the two resonances: stated differently, the loss of the resonance depends on the resonance frequency of $a$. Physically, the resonance-frequency-dependence of the loss arises from interference: light from the resonator $a$ can decay by directly leaking into the continuum, or by hopping through the resonator $d$ first. These two pathways interfere, and their relative phase depends on the frequency of $a$. 

This model has been studied by many authors for the key feature that it supports conditions where the resonances can be lossless, despite both resonators being coupled to the continuum. These lossless states are often referred to as bound states in the continuum (BICs). Contrary to standard works \cite{hsu2016bound, azzam2021photonic, marinica2008bound, fan2002analysis}, we consider the case in which one of the resonances is nonlinear, and consider the quantum optical consequences of this modification.

For the F.W. model, the coupling and loss functions for the resonance $a$ are:
\begin{subequations}
\begin{align}
    K_c(\omega) &= \sqrt{2\kappa}  \left[1 - \frac{\gamma}{i(\omega_d - \omega) + \gamma}\right] \label{eq:fw_Kc} \\
    K_l(\omega) &= \kappa  \left[1 - \frac{\gamma}{i(\omega_d - \omega) + \gamma}\right], \label{eq:fw_Kl}
\end{align}
\end{subequations}
where $\kappa$ and $\gamma$ are the respective decay rates of $a$ and $d$, which have frequencies $\omega_{a,d}$. 
The model's key feature is that interference between radiative channels equips $a$ with a resonance-frequency-dependent loss rate that vanishes at $\omega_d$ (Fig.~\ref{fig:mean_field}b). 

\subsection{Mean-field dynamics}


Non-Markovian environment coupling strongly impacts the classical (mean-field) dynamics of $a$. These dynamics are governed by Eq.~\ref{eq:general_eom}, taking all operators to c-numbers to yield a generalized TCMT model with non-Markovian loss. When light is injected into the cavity, the input fields act as source terms. For a monochromatic pump of frequency $\omega_p$, the input field has mean value $\braket{s(t)} = s_0 e^{-i\omega_p t}$, where $|s_0|^2$ is the incident flux.
In this case, the cavity photon number $n = \braket{a^\dagger a} \approx |\braket{a}|^2$ in the steady-state satisfies the cubic equation:
\begin{equation}
    [(\omega_{ap} + K_l''(\omega_p) + \beta n)^2 + K_l'(\omega_p)^2]n = |s_0|^2 |K_c(\omega_p)|^2,
    \label{eq:bistab_steady_state}
\end{equation}
where, $\omega_{ap} \equiv \omega_a - \omega_p$, and $K_l = K_l' + iK_l''$ has been decomposed into real and imaginary parts which respectively give the dispersive loss and phase shift. The right hand side concerns the coupling of the input, while the left hand side concerns the cavity response, which is sensitive to the loss and detuning at the pump frequency.

As an important point of comparison to existing literature, we note that an externally pumped cavity with Kerr nonlinearity can exhibit bistability \cite{drummond1980quantum}. In other words, for a given pump strength $s_0$, there can be two stable solutions to Eq.~\ref{eq:bistab_steady_state}. This occurs because the amount of light coupled into the cavity depends on the pump-cavity detuning, which in turn depends on the cavity photon number via Kerr nonlinearity. 
Bistability occurs when $\delta > \sqrt{3} K_l'(\omega_p)$, where $\delta \equiv \omega_{ap} + K_l''(\omega_{ap})$ is the total detuning (see dashed curves in Fig.~\ref{fig:mean_field}c).


In the non-Markovian case, the cavity exhibits a different coupling into the cavity for each pump frequency, and a corresponding different loss. As a result, the non-Markovian input-output behaviors exhibit important deviations from their Markovian counterparts.
The most important distinction occurs when the pump frequency is near the frequency of the lossless mode (BIC), which is approximately $\omega_d$ when $\kappa \ll \gamma$. 
Here, the F.W. model exhibits vanishing loss, which drastically reduces the pump power required to maintain the steady-state at the boundary of the top stable branch. In particular, it can be shown through Eq.~\ref{eq:bistab_steady_state} that at the upper bistable point, the required pump flux is reduced by the same proportion that the loss is reduced (2 orders of magnitude in this example).
The combination of Kerr nonlinearity and a frequency for which the loss nearly vanishes thus corresponds to a cavity intensity for which the loss nearly vanishes. Although low threshold bistability can of course be realized in a Markovian system (by minimizing the loss), this loss rate will still be independent of the cavity intensity.

Non-Markovian dissipation also strongly impacts the transient behaviors in nonlinear systems, in particular by introducing a modulational instability (MI) which occurs specifically due to the frequency-dependence of the loss.
By analyzing the response of the steady-state to perturbations, we construct the phase diagram of the driven-dissipative system in the space of pump frequency and photon number (Fig.~\ref{fig:mean_field}d). The black region indicates the traditional unstable region which occurs when the input-output curve ``bends back'' on itself, splitting the input-output curve into two disconnected stable branches.  
The grey regions mark the MI induced by loss dispersion. In this particular example, the MI gain is highest in the vicinity of the bistable region (see S.I.). 

The consequences of this modulation instability can be understood from the transient dynamics of the nonlinear system excited from the vacuum state with various pump fluxes. For parameters in the stable part of the phase diagram, transients can be described as a damped oscillation around the steady-state (referred to sometimes as ``relaxation oscillations'') (Figs.~\ref{fig:mean_field}e-f). For a pump flux which nears, but does not enter, the MI regime (Fig.~\ref{fig:mean_field}f), the decay time of the relaxation oscillations increases by an order of magnitude, compared to a point further from the instability (Fig.~\ref{fig:mean_field}e).

Inside the MI region (Fig.~\ref{fig:mean_field}g), the photon number pulses about the steady-state value predicted from the mean-field theory ($n=5\times 10^8$), spontaneously and indefinitely. The pulsing frequency is $\Omega_p = \sqrt{\Omega^2 - (\Gamma/2)^2}$, where $\Omega = \sqrt{\Delta^2 - (\beta n)^2}$ is the relaxation oscillation frequency given in terms of the detuning parameter $\Delta = \omega_{ap} + 2\beta n$, and $\Gamma$ is the relaxation oscillation decay rate. 
The pulsing amplitude can be a substantial fraction of the steady-state value; for these particular parameters, the cavity photon number swings over a range which is 80\% of the mean value itself.

The MI which occurs here is initiated not by frequency dispersion of the index of refraction \cite{tai1986observation}, but rather the dispersion of the loss of the cavity. A small number of works have explored MI induced by dispersive loss in optical fibers \cite{tanemura2004modulational, perego2018gain, bessin2019gain}. Although the MI in the system described here shares the dispersive loss feature with known fiber systems, the cavity nature of our system results in some key differences in the way the pulsing frequency $\Omega_p$ is set. In particular, the MI we describe has a pulsing frequency which depends on the steady-state photon number, in contrast to the related fiber effects, in which the MI frequency is set by the detuning of the pump from the maximum loss frequency. These differences make it clear that the physical nature of the instability here is substantially different than previous works in fibers, even though disperisve loss is needed in both cases. For the parameters considered here, $\Omega_p$ is on the THz scale, giving such sources an attractive potential to modulate light at frequencies which are inaccessible by electronic means.

\begin{figure*}
    \centering
    \includegraphics{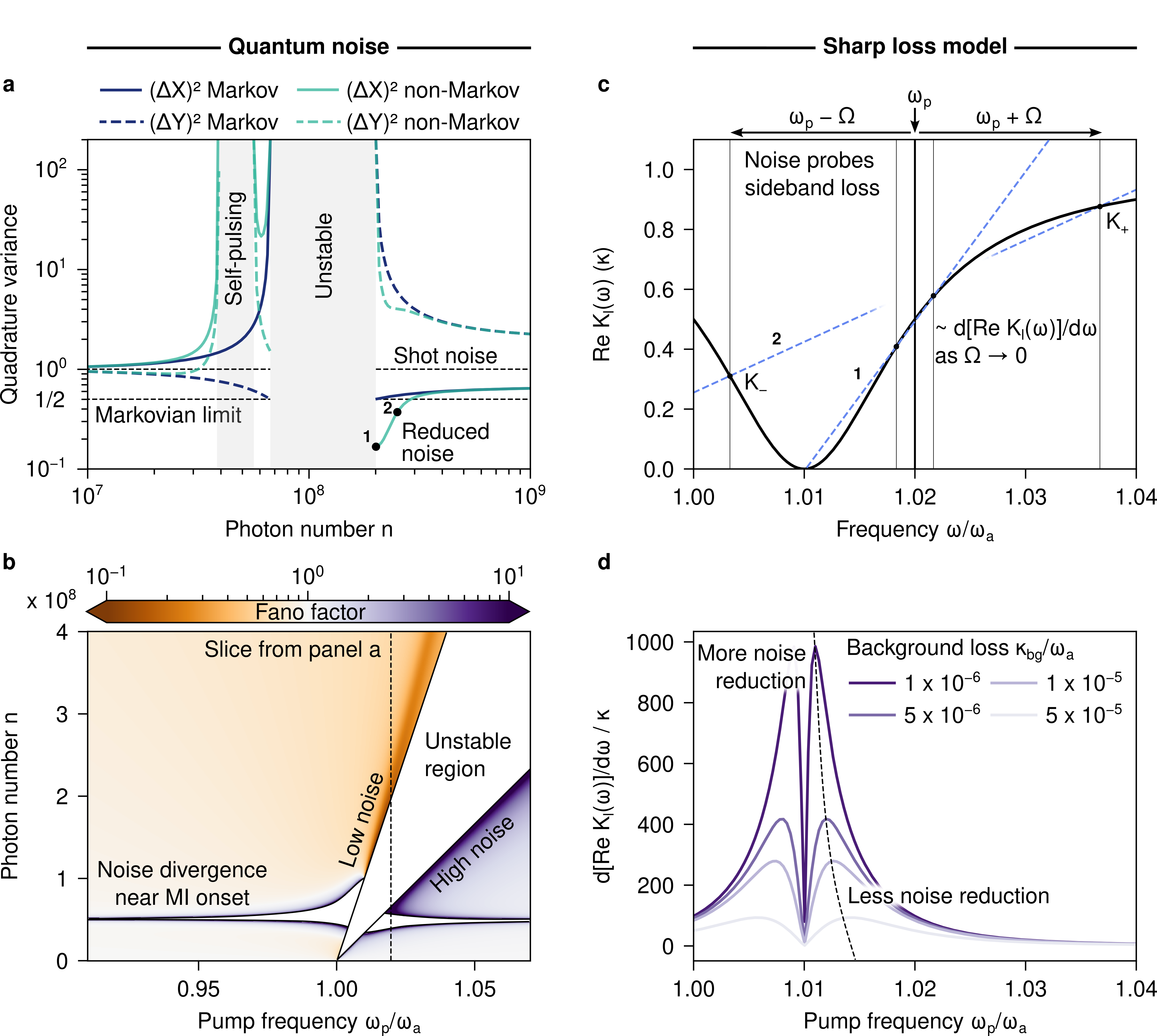}
    \caption{\textbf{Quantum noise dynamics in a non-Markovian driven-dissipative cavity.} (a) Amplitude (X) and phase (Y) quadrature noise of the nonlinear F.W. model as a function of the steady-state photon number $n$. Quadrature variance is shown in reference to the shot noise level. The non-Markovian model exhibits significant noise reduction on the upper bistable branch. The pump frequency is $\omega_a = \omega_d + \gamma$, and all system parameters are the same as those used in Fig.~\ref{fig:mean_field}. (b) Nonlinear phase diagram which shows amplitude quadrature variance as a function of pump frequency $\omega_p$ and steady-state photon number $n$. The Fano factor is defined as $F = (\Delta n)^2/n$. (c) Sharp loss interpretation of amplitude noise reduction based on the dispersive loss. Pump frequency and sidebands at $\pm\Omega$ probe the contours of the loss. As photon number approaches the bistable point, the relaxation oscillation frequency $\Omega$ approaches zero, so that the noise probes the derivative of the loss. (d) Sharpness of loss as a function of different background losses.}
    \label{fig:quantum_noise}
\end{figure*}

\subsection{Quantum noise dynamics}

In addition to the mean-field properties described above, non-Markovian loss in driven-dissipative systems can induce important changes in quantum noise properties. Namely, amplitude and phase noise increase near the onset of MI, while amplitude noise can be strongly suppressed below the shot noise limit by the presence of sharp resonance-frequency-dependent loss.

This discussion is based on our theory for the quantum noise of the cavity in the presence of nonlinearity and non-Markovian dissipation. The theory is based on a linearization approach that considers deviations of Eq.~\ref{eq:general_eom} from the mean-field solution. In particular, we assume that in the steady-state, the cavity annihilation operator can be decomposed into a mean value and operator-valued fluctuations as $a = \braket{a} + \delta a(t)$. By substituting this expression into Eq.~\ref{eq:general_eom}, maintaining contributions at linear order, and transforming into frequency space, we obtain a coupled set of equations for the frequency space noise operators:
\begin{equation}
    \begin{pmatrix}
        \eta(\omega) & i\beta n \\
        -i\beta n & \eta^*(-\omega)
    \end{pmatrix} \begin{pmatrix}
        \delta a(\omega) \\ \delta a^\dagger(-\omega) 
    \end{pmatrix} = \begin{pmatrix}
        K_c(\omega_p + \omega) \delta s(\omega) \\ K_c^*(\omega_p - \omega) \delta s^\dagger (-\omega)
    \end{pmatrix}.
\end{equation}
Here, we have defined $\eta(\omega) \equiv i(\omega_a - \omega_p - \omega + 2\beta n) + K_l(\omega_p + \omega)$. Additionally, it it is understood that $n = \braket{a^\dagger a}$ refers to the steady-state mean-field value.



The noise properties are most compactly described in terms of the variances of quadrature operators $X = a + a^\dagger$ and $Y = -i(a - a^\dagger)$, which we find to be: 
\begin{equation}
    (\Delta Q_\sigma)^2 = \int \frac{d\omega}{\pi} \frac{R_\sigma(\omega)}{\left[\Omega^2(\omega) - \omega^2\right]^2 + \Gamma^4(\omega)},
    \label{eq:quad_noise_exact}
\end{equation}
\begin{equation}
    R_\sigma(\omega) \equiv K_+(\omega)\left[(\omega_{ap} + (2 - \sigma)\beta n + \omega)^2 + K_-(\omega)^2\right].    
\end{equation}
Here, $K_\pm(\omega) \equiv \text{Re}\,K_l(\omega_p \pm \omega)$, and $\sigma$ denotes the quadrature with variance $(\Delta Q_\sigma)^2$, with $\sigma = 1$ corresponding to $X$ and $\sigma=-1$ corresponding to $Y$. Additionally, $\Omega(\omega)$ and $\Gamma(\omega)$ are the frequency-dependent relaxation oscillation frequency and decay rate (see S.I. for expressions). In this formulation, the phase of the pump is chosen, without loss of generality, so the mean-field steady-state $\braket{a}$ is positive and real; then, $\Delta X$ is the intensity noise, and $\Delta Y$ the phase noise.

As a point of comparison, we briefly review the noise properties associated with an ordinary bistable cavity (Fig.~\ref{fig:quantum_noise}a, Markov). On the lower bistable branch, the amplitude noise increases with photon number until eventually diverging at the lower bistable point, 
while, the phase noise decreases with photon number until hitting a minimum value which lies a factor of 2 below the shot noise limit (SNL). On the upper bistable branch, the situation is reversed: the phase noise diverges near the bistable point, while the amplitude noise attains a minimum value of $1/2$ relative to the SNL. At large photon numbers, the amplitude noise approaches a universal value of $2/3$ of the SNL for all pump frequencies \cite{drummond1980quantum}. 

Dispersive dissipation introduces dramatic changes.
First, both amplitude and phase noise diverge near the onset of the MI, indicating the presence of strong bunching \footnote{Values for noise are not shown within the self-pulsing region since the system does not reach a steady-state, and thus a notion of steady-state quantum noise is not defined.}. 
Second, the amplitude noise at the onset of the upper bistable branch drops dramatically compared to the Markovian case. This region lends itself to the natural generation of highly intensity-squeezed states; for the set of parameters shown, the minimum amplitude variance reached is around 0.15, almost 10 dB below the shot noise limit. By tuning the detuning of $\omega_a$ and $\omega_d$, as well as the pump $\omega_p$, it is even possible to exceed 10 dB of amplitude squeezing (see S.I.). 

\begin{figure}
    \centering
    \includegraphics{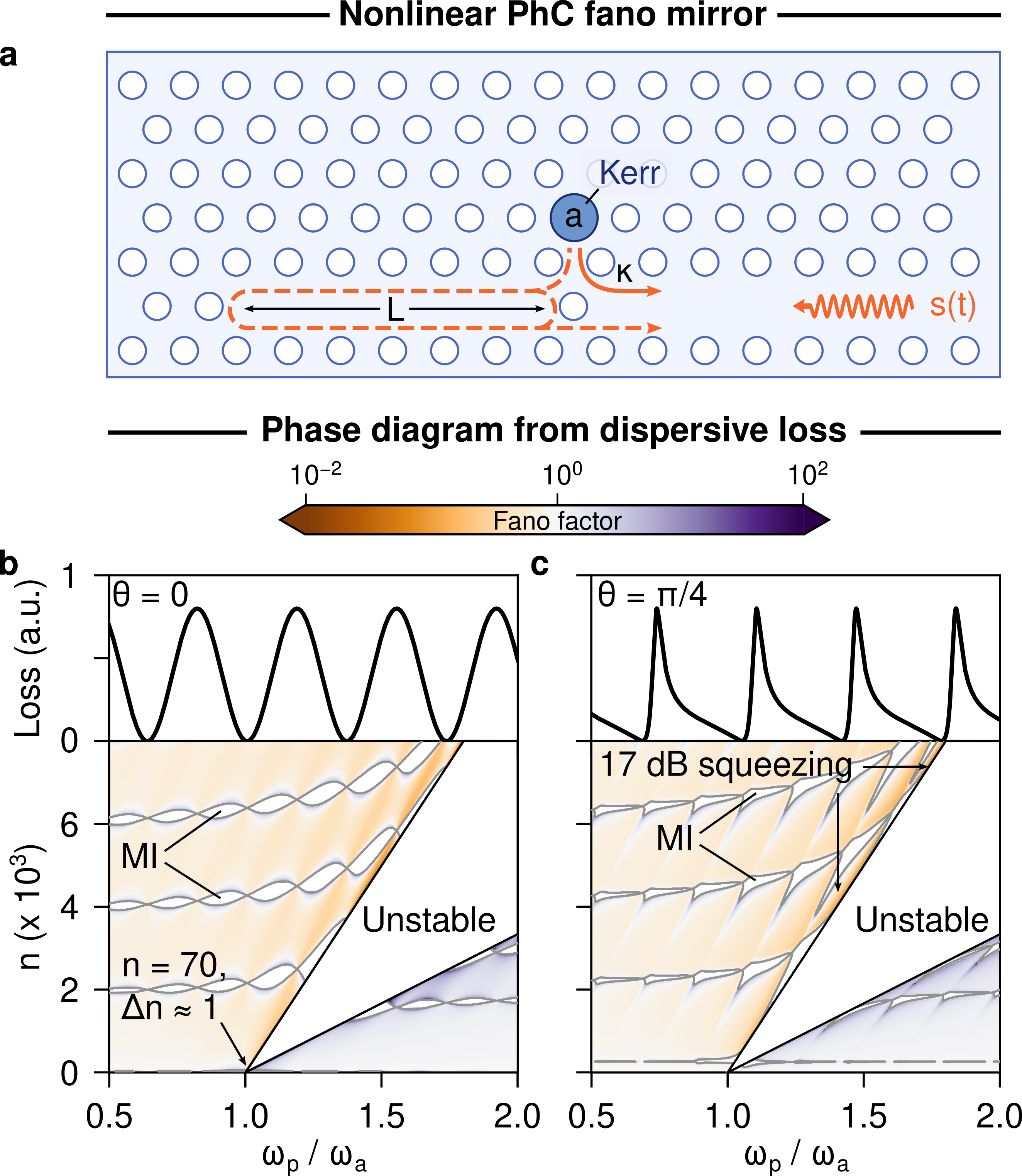}
    \caption{\textbf{Driven-dissipative behavior of a nonlinear Fano mirror.} (a) Photonic crystal (PhC) based Fano mirror which exhibits dispersive loss as a result of interference between a line cavity and a single defect cavity. Parameters used are $\omega_a = 1.03 \times 10^{15}$ s$^{-1}$, $\kappa = 10^{-4} \omega_a$, $L = 5$ $\mu$m, and $\beta = 10^{-4}\omega_a$, , which are chosen to represent systems with strong per-photon nonlinearity \cite{fink2018signatures}. (b, c) Loss profile of the resonator $a$ for different values of the reflection and transmission coefficients at the junction between the resonances and the external continuum. Profiles can be either be symmetric (b) or asymmetric (c). Below each loss profile is the phase diagram of the system subject to continuous driving with frequency $\omega_p$ to a steady state photon number $n$. The ``wedge'' region is unstable through the traditional mechanism, while the many other regions exhibit modulational instability.}
    \label{fig:fano}
\end{figure}

This physics behind this phenomenon is best understood through the lens of intensity-dependent dissipation \cite{rivera2023creating}. 
In particular, the simultaneous presence of Kerr nonlinearity and strong resonance-frequency-dependent loss provides the cavity with an effective \emph{photon number dependent loss}, obtained by composing the resonance-frequency-dependent loss and the intensity-dependent resonance frequency. In the quantum picture, certain number states experience much lower losses than others, effectively amplifying their presence in the steady-state. By incorporating this idea of intensity-dependent loss with a continuous-wave drive, it then becomes possible to indefinitely stabilize low-noise resonator states, and even states approaching intracavity Fock states (Fig.~\ref{fig:fano}). That such low-noise resource states can be indefinitely maintained --- even in the presence of loss --- shows a critical advantage of considering non-Markovian nonlinear dynamics in the driven-dissipative setting.  

This intuition is well supported by an analytical approximation we have derived for the amplitude noise (see S.I.):
\begin{equation}
    (\Delta X)^2 \approx \left(1 - \frac{\beta n}{\Delta}\right)\frac{\left(\frac{\Delta}{\Omega}\right)^2 + r\left(\frac{\Delta}{\Omega}\right)}{1 + r\left(\frac{\Delta}{\Omega}\right)}.
    \label{eq:noise_approx}
\end{equation}
This approximation holds in the adiabatic regime in which the non-Markovian coupling element has a shorter time response than cavity ($\gamma \gg \kappa$ for the F.W. model). In Eq.~\ref{eq:noise_approx}, $r = (K_+ - K_-)/(K_+ + K_-)$ is the ratio of the difference and sum of losses at sideband frequencies $K_\pm \equiv \text{Re}\,K_l(\omega_p \pm \Omega)$. Thus, while the steady-state mean-field behavior depends on the loss at the pump frequency (Eq.~\ref{eq:bistab_steady_state}), the noise level is determined by the difference in loss between the sideband frequencies $\omega_p \pm \Omega$ (Fig.~\ref{fig:quantum_noise}c). A sharply resonance-frequency-dependent loss creates a nonzero $r$, allowing the Fano factor to drop far below the Markovian limit of $1/2$. As the upper bistability boundary is approached, the sideband frequency $\Omega$ tends to zero, so that $r$ depends on the slope of the dispersive loss at the pump frequency. In particular, $r/\Omega \to [d\,K_l'(\omega)/d\omega]_{\omega_p}/K_l'(\omega_p)$ as $\Omega \to 0$. In this regime, the amplitude noise is determined quite directly by the ``sharpness'' of the dispersive loss curve: a sharper loss pushes noise further below the classical limit.

One critical question is the maximum amplitude noise reduction below the SNL which can be achieved through coherent pumping of a nonlinear non-Markovian resonance. Unlike other methods for generating amplitude-squeezed light with Kerr nonlinearity \cite{bondurant1984squeezed, kitagawa1986number}, the approach presented here does not have strict theoretical limits on the achievable number uncertainty. This is possible since the ideal noise reduction is set by the ratio of the sharpness of the frequency-dependent loss to the loss itself. For the F.W. model, this can be achieved by bringing the pump frequency to an $\omega_p$ arbitrarily close to the BIC at $\omega_d$. In practice, the limiting factor on the noise reduction will be the presence of additional loss mechanisms such as material absorption, scattering loss, or multi-photon decay processes \cite{rivera2023creating}. To model this, we add a small frequency-independent background loss $\kappa_{\text{bg}}$ to the resonator $a$. 
Larger background losses decrease the maximum ratio $r$ that can be achieved, and push the pump frequency of maximal noise reduction slightly away from the BIC (Fig.~\ref{fig:quantum_noise}d).

\emph{More general dispersive loss models.} Although we have focused on the nonlinear F.W. model as a minimal example, both the physical phenomena and theoretical tools in this work extend much more widely.
As an example, we briefly show that similar phenomena can be realized in a reflection-based Fano resonance geometry with strong nonlinearity. Such structures have been realized experimentally, and even shown to exhibit classical nonlinear effects \cite{yang2020inverse, van2022all, yu2021ultra, tanaka2007dynamic}, making them prime candidates for quantum optics experiments. For the particular example we consider (Fig.~\ref{fig:fano}a), the loss is periodic in frequency, with period set by the free spectral range of the cavity. The loss profile can be engineered by tuning the transmission and reflection parameters of the Fano interference (Figs.~\ref{fig:fano}b-c). 

By pumping to photon numbers near the upper bistable onset, one can create cavity states with amplitude noise far below the classical limit.
Thus the ``sharp loss'' mechanism of noise reduction is quite general, and can be exploited in many systems with sufficiently sharp dispersion, strong nonlinearity, and low background loss. For this particular system, pumping at a frequency which is highly detuned from $\omega_a$ enables the generation of states which contain several thousand photons, with Fano factors down to $0.02$ (equivalent to 17 dB of squeezing). Moreover, small detunings enable the generation of states which become very close to intracavity Fock states: this particular example system supports a cavity state containing $n=70$ photons, with an uncertainty $\Delta n \approx 1$ (Fig.~\ref{fig:fano}b).

The Fano mirror loss also introduces additional complexities in the phase diagram and quantum noise landscape compared to the F.W. model (Figs.~\ref{fig:fano}b-c). 
Since the stability behavior and quantum noise are sensitive to the sideband losses at $\omega_p \pm \Omega$, varying the pump frequency and steady-state number probes the dispersive loss. The result is that both the quantum noise and MI regions produce a quilted pattern, with MI boundaries determined by the zeros of the denominator in Eq.~\ref{eq:noise_approx}. By changing the dispersion of the loss (in this case, by changing the reflection and transmission coefficients at of the Fano interference), one changes the instability pattern.

\emph{Experimental considerations.} The architectures and parameters used in this work were chosen for their compatibility with current experimental capabilities. In particular, the parameters used in Figs.~\ref{fig:mean_field} and \ref{fig:quantum_noise} are compatible with state of the art integrated nanophotonic structures \cite{van2022all, yang2020inverse}. Such structures, as well as other potential structures on silicon nitride or lithium niobate, are appealing due to their simultaneous ability to provide highly optimized classical interference for dispersive dissipation, and substantial Kerr nonlinearity with a compact footprint. Additionally, exciton-polariton condensates in nanophotonic structures \cite{fink2018signatures}, can provide the modal confinement and nonlinear strength required to realize these effects with tens to thousands of photons (as shown in Fig.~\ref{fig:fano}). Yet other more ``macroscopic'' platforms may also prove suitable hosts for the non-Markovian driven-dissipative platforms we describe; fiber systems in particular have the advantage of long propagation lengths to provide the requisite nonlinear phase shifts.

\section{Conclusion and outlook}

In summary, we have introduced non-Markovian driven-dissipative systems as a platform to engineer the classical and quantum behavior of light. In these systems, the dependence of the loss rate on resonance frequency leads to driven-dissipative phases which are not present with Markovian dissipation. In particular, we showed that dispersive dissipation in coherently driven cavities leads to an unexplored class of modulational instability. Additionally, we showed that in these systems, the amplitude noise is shaped by the dispersion of the loss, with sharp resonance-frequency-dependent losses leading to the natural steady-state production of strongly intensity squeezed, and even near-Fock cavity states.


Unlike most driven-dissipative systems which interact instantaneously with their environments, non-Markovian architectures harness temporal correlations to greatly enlarge the space of possible behaviors.  
Our findings point toward the potential of these systems to create quantum states of light which are not naturally produced by other means. In addition to the strong intensity squeezing described, using these platforms to generate multi-mode correlated states could open new avenues in optical quantum information processing. 

Although we have focused on Kerr nonlinearity in this work, we anticipate that it will be fruitful to explore more general non-Markovian driven-dissipative systems. In particular, by incorporating phase-sensitive dynamics (such as those realized by optical parametric oscillators and amplifiers \cite{nehra2022few, roy2023non}), an even more interesting range of behaviors is expected. Moreover, the incorporation of gain media into nonlinear cavities can provide yet a another route to realizing strong non-Markovian effects \cite{pontula2022strong, nguyen2023intense}. Finally, we note that while we considered continuous-wave driving, the physics here can also be extended to generating quantum states with pulsed driving, an ongoing topic of theoretical and experimental exploration \cite{yanagimoto2022onset, guidry2022quantum, ng2023quantum, guidry2023multimode}. These platforms --- and others not yet imagined --- have the potential to bring new degrees of control to the classical and quantum behavior of light. 

\section{Acknowledgments}

We acknowledge useful discussions with Yannick Salamin and Shiekh Zia Uddin. J.S. acknowledges previous support of a Mathworks Fellowship, as well as previous support from a National Defense Science and Engineering Graduate (NDSEG) Fellowship (F-1730184536). N.R. acknowledges the support of a Junior Fellowship from the Harvard Society of Fellows. This work is also supported in part by the U. S. Army Research Office through the Institute for Soldier Nanotechnologies at MIT, under Collaborative Agreement Number W911NF-18-2-0048. We also acknowledge support of Parviz Tayebati.

\bibliographystyle{unsrt}
\bibliography{main.bib}

\end{document}


\rmfamily

\title{Supplementary information for: \\
Driven-dissipative phases and dynamics in nonlinear non-Markovian photonics}
\author{Jamison Sloan$^{1,\dagger}$, Nicholas Rivera$^{2,3,\dagger}$, and Marin Solja\v{c}i\'{c}$^{1,3}$}

\affiliation{$^{1}$Research Laboratory of Electronics, Massachusetts Institute of Technology, Cambridge, MA 02139, USA. \\
$^{2}$Department of Physics, Harvard University, Cambridge, MA 02138, USA.  \\
$^{3}$Department of Physics, Massachusetts Institute of Technology, Cambridge, MA 02139, USA}
\noindent	

\noindent

\clearpage

\renewcommand{\sp}{\sigma_+}
\newcommand{\sm}{\sigma_-}

\setlength{\parindent}{0em}
\setlength{\parskip}{.5em}
\vspace*{-2em}


\newcommand{\bin}{b_{\text{in}}}
\newcommand{\bbarin}{\bar{b}_{\text{in}}}
\begin{abstract}
    In this Supplementary Information (S.I.), we derive the results of the main text, related to dispersive coupled resonances in the presence of intensity-dependent nonlinearity. We use the input-output formalism to show how these systems can be used to generate quantum states of light when coherently pumped. We also perform stability analysis to show how these systems can exhibit modulational instability due to non-Markovian dissipation.
\end{abstract}

\maketitle

\tableofcontents

\newpage

\section{General formalism for non-Markovian dissipation}

We begin by developing the general formalism for the description of a nonlinear cavity mode with non-Markovian loss \cite{rivera2023creating}. Such a non-Markovian loss arises from coupling to frequency-dependent reservoirs, and result in frequency-dependent dissipation. In the presence of intensity-dependent (Kerr) nonlinearity, such frequency-dependent dissipation can lead to an effective intensity-dependent dissipation. 

\subsection{Heisenberg-Langevin equations of motion}
The equation of motion for the annihilation operator of a Kerr nonlinear resonance $a$ coupled to non-Markovian dissipation channels $i$ is:
\begin{equation}
    \dot{a} = -i\omega_a a - i\beta(a^\dagger a) a - \int dt'\,K_l(t-t')a(t') + \sum_i \int dt'\, K_{c,i}(t-t')s_i(t').
    \label{eq:general_eom}
\end{equation}
Here, $\omega_a$ is the bare frequency of the resonance $a$, $\beta$ is the Kerr frequency shift per photon, $K_l(t-t')$ is a time-domain loss Kernel which provides frequency-dependent damping of $a$ through a convolution, and $K_{c,i}(t-t')$ describe the coupling of external input fields $s_i(t)$ in through various channels $i$. Each $i$ can be used to describe the coupling of the mode $a$ to a different reservoir. In the frequency-domain, the coupling and loss kernels are associated with frequency-dependent functions
\begin{equation}
    K_{c,l}(\omega) = \int dt\, e^{i\omega t} K_{c,l}(t).
\end{equation}
These functions can be constructed through theoretical modeling (see Section~\ref{sec:loss_models}), or empirically through experimental data.

Additionally, we note that the input fields are normalized so that $[s_i(t), s_j^\dagger(t')] = \delta_{ij}\delta(t-t')$, which means that $|s_i(t)|^2$ has units of photon flux in photons/s. Going forward, we will sometimes omit the reservoir index $i$ from the input fields and coupling kernels if it is assumed that there is only one input channel.

\textbf{Markovian limit:} We briefly describe how the theory reduces to the more typically used Markovian limit. Physically, the Markovian limit corresponds to situations where the bandwidth of the reservoir is large compared to that of the mode under consideration, resulting in a so-called ``white noise'' coupling. In the time-domain, this is equivalent to the condition that the cavity mode and reservoir interact instantaneously: any internal dynamics of the coupling are fast compared to all other relevant timescales. In this case, the cavity is described by a frequency independent loss rate $\gamma$. The loss rate determines the coupling and loss functions as $K_l(\omega) = \gamma$, and $K_c(\omega) = \sqrt{2\gamma}$. In the time-domain, the loss and coupling functions are delta functions (indicating the instantaneous response), so that for a single Markovian coupling channel, one recovers the equation of motion:
\begin{equation}
    \dot{a} = -i\omega_a a - \gamma a + \sqrt{2\gamma} s(t).
\end{equation} 
The quantum fluctuations of the source $s$ generate the Langevin force $F_\gamma(t) = \sqrt{2\gamma}\delta s(t)$. In particular, we have decomposed the input field as $s(t) = \braket{s(t)} + \delta s(t)$. For a vacuum reservoir, $\braket{F_\gamma(t)F_\gamma^\dagger(t')} = 2\gamma \delta(t-t')$, while all other quadratic expectation values of $F_\gamma$ and $F_\gamma^\dagger$ vanish. 

\textbf{Classical limit:} In the classical limit, defined by the expectation value  $\braket{\cdot}$ which converts operators to c-numbers, Eq.~\ref{eq:general_eom} reduces to a temporal coupled mode theory (TCMT) equation which includes the time delay effects needed to describe memory effects. In this case, we obtain the classical non-Markovian TCMT equation:
\begin{equation}
    \dot{\alpha} = -i\omega_a \alpha - i\beta|\alpha|^2 \alpha - \int dt'\,K_l(t-t')\alpha(t') + \sum_i \int dt'\, K_{c,i}(t-t')\braket{s_i(t')}.
    \label{eq:general_eom_tcmt}
\end{equation}
Here, we have used the notation $\alpha = \braket{a}$, and the sources $\braket{s_i(t)}$ are now c-number classical sources. Due to the nonlinearity, this TCMT equation makes the additional approximation that $\braket{a^\dagger a a} \approx \braket{a^\dagger}\braket{a}\braket{a} = |\alpha|^2\alpha$. Such an approximation provides an accurate description of systems with quantum states which are localized in phase space, or in other words, states which are well-described as a coherent part, with small added quantum fluctuations. An equation such as Eq.~\ref{eq:general_eom_tcmt} admits numerical solution with split-step methods, in which the Kerr nonlinearity is treated in the time domain (due to the time-local nature of the term), and the non-Markovian terms are treated in the frequency domain (where the convolutions become products such as $K_l(\omega)a(\omega)$).

If the Markovian and classical limits are taken together, we recover the standards temporal coupled mode theory expression for a leaky resonance:
\begin{equation}
    \dot{\alpha} = -i\omega_a \alpha - \gamma \alpha + \sqrt{2\gamma}\braket{s(t)}.
\end{equation}

\subsection{Kramers-Kronig relations and non-Markovian correlation functions}

The coupling and loss kernels are not independent from one another. This is a manifestation of the fact that the frequency-dependent damping rate arises from the frequency-dependent coupling of the resonance $a$ to the external reservoirs. In fact, the knowledge of one of these functions is sufficient to determine the other. The loss and coupling functions are related in the frequency-domain by a Kramers-Kronig relation:
\begin{equation}
    K_{l}(\omega) = i\int\frac{d\omega'}{2\pi} \frac{|K_{c}(\omega')|^2}{\omega-\omega'+i\eta},
\end{equation}
with $\eta$ infinitesimal. This is consistent with a non-Markovian generalization of the so-called ``Einstein relations'' which link drift and diffusion terms in Heisenberg-Lanegvin equations in Markovian settings \cite{scully1999quantum}.

We will describe some additional properties of the non-Markovian correlation functions. To do so, we will consider Eq.~\ref{eq:general_eom} without any Kerr nonlinearity $(\beta = 0)$. Additionally, we will assume that the input field $s(t)$ has no mean value, and consists only of operator-valued vacuum fluctuation $\delta s(t)$. Physically, this corresponds to the case where light is allowed to leak out of the cavity from an initial state, without additional energy input. Such an assumption allows us to isolate the action of the non-Markovian reservoir on the cavity, and enables us to write the equation of motion as: 
\begin{equation}
    \dot{a} = -i\omega_a a - \int dt'\, K_l(t-t')a(t') + F(t).
    \label{eq:non-markov_langevin}
\end{equation}
In this expression, the final term is a Langevin force defined as 
\begin{equation}
    F(t) \equiv \int dt'\, K_c(t-t') \delta s(t').
\end{equation}
In the time domain, the correlator of the Langevin force is given as 
\begin{equation}
    \braket{F(t) F^\dagger(t')} = \int dt''\, K_c(t - t'') K_c^*(t' - t'').
\end{equation}
We thus immediately see that this relation will in general not be a delta function. 

This result can also be used to obtain correlations in the frequency domain, which have a much more clear interpretation. In the frequency domain, this input field commutation relations are $[s(\omega), s^\dagger(\omega')] = 2\pi\delta(\omega-\omega')$. Using this, as well as the time domain correlator for $F$ shown above, the nonzero correlator of the frequency domain Langevin force $F(\omega)$ can be evaluated as
\begin{align}
    \braket{F(\omega) F^\dagger(\omega')} &= \int dt\,dt'\, e^{i\omega t}e^{-i\omega't'} \braket{F(t) F^\dagger(t')} \\
        &= \int dt\,dt'\,dt''\, e^{i\omega t}e^{-i\omega't'} K_c(t-t'')K_c^*(t'-t'') \\
        &= 2\pi |K_c(\omega)|^2 \delta(\omega - \omega').
\end{align}
As noted earlier, the modulus squared of the coupling function is proportional to the frequency-dependent loss $K_l'(\omega)$, and thus the correlator can also be written as $\braket{F(\omega) F^\dagger(\omega')} = 2\,\text{Re}\,K_l(\omega) \cdot 2\pi\delta(\omega-\omega').$ Thus, we arrive at the intuitive conclusion that the frequency domain correlations are proportional to the frequency-dependent loss $\text{Re}\,K_l(\omega)$ that results from the frequency-dependent out-coupling. Moreover, the correlator still retains a delta function in the frequency domain, indicating that fluctuations at different frequencies remain uncorrelated. This is the case because even though the system we describe is dispersive, time-translation invariance is still retained. 

\subsection{Preservation of commutation relations}

Now, we will prove that the model used to describe the non-Markovian damping of the resonance $a$ preserves the commutation relations $[a,a^\dagger] = 1$, which is required for any quantum-mechanically consistent theory of dissipation. We will continue to neglect nonlinearity: it can be shown that if the equation without $\beta$ conserves the commutator, then the equation with nonzero $\beta$ does as well. Since the operator equation of motion for $a$ (Eq.~\ref{eq:non-markov_langevin}) is linear, it can be transformed into the frequency domain without approximation. By doing this, we find that $a(\omega)$ is related to the Langevin force as 
\begin{equation}
    a(\omega) = \frac{F(\omega)}{-i(\omega_a - \omega) + K_l(\omega)} \equiv i\xi(\omega)F(\omega).
\end{equation}
As defined, $\xi(\omega)$ is a response function which governs how the input field (and its fluctuations) couple into the resonance $a$ as a function of frequency. We can then express the equal time commutation relation of $a$ as
\begin{align}
    [a(t) , a^\dagger(t)] &= \int \frac{d\omega}{2\pi}\frac{d\omega'}{2\pi} e^{-i\omega t}e^{i\omega' t} [a(\omega), a^\dagger(\omega')] \\
    &= \int\frac{d\omega}{2\pi}\frac{d\omega'}{2\pi} e^{-i\omega t}e^{i\omega' t} \xi(\omega)\xi^*(\omega') [F(\omega), F^\dagger(\omega')] \\
    &= \int \frac{d\omega}{\pi} |\xi(\omega)|^2 \mathrm{Re}\,K_l(\omega).
\end{align}
We note that in the final step, we have taken an expectation value of the Langevin force commutator with respect to the reservoir. We see then that the commutation relation of $a$ is expressed as a frequency-dependent integral containing the frequency-dependent loss rate, as well as the response function $\xi(\omega)$ which describes the linear coupling of the input field into the cavity. 

Now, we will show that this integral can be written in a simpler and more fundamental form, using the analytic structure of $\xi$. In particular, we note that 
\begin{equation}
    |\xi(\omega)|^2 = \frac{1}{(\omega_a + K_l''(\omega) - \omega)^2 + K_l'(\omega)^2},
\end{equation}
where for brevity we will sometimes express the real and imaginary parts of $K_l$ as $K_l = K_l' + iK_l''$. We then note that the imaginary part of $\xi$ is given as
\begin{equation}
    \text{Im}\,\xi(\omega) = \frac{K_l'(\omega)}{(\omega_a + K_l''(\omega) - \omega)^2 + K_l'(\omega)^2}.
\end{equation}
Thus, the commutator integrand can be written as 
\begin{equation}
    [a(t) , a^\dagger(t)] = \int \frac{d\omega}{\pi} |\xi(\omega)|^2 \mathrm{Re}K_l(\omega) = \int \frac{d\omega}{\pi} \text{Im}\, \xi(\omega).
    \label{eq:commutator_integral}
\end{equation}
By using this form, the commutation relation is written as a frequency integral over the imaginary part of a single function.

To conclude the proof, we will show that $\xi$ is a response function which obeys its own Kramers-Kronig (K.K.) relation, and develop a sum rule which evaluates the integral of Eq.~\ref{eq:commutator_integral}. Recall that $\xi(\omega)$ is defined such that in the absence of any input field, $a(\omega) = -i\xi(\omega) F(\omega)$. This frequency-domain multiplicative relation corresponds to a time-domain convolution $a(t) = -i\int dt'\,\xi(t-t')F(t')$. Since this system is linear and time invariant, the time domain response function $\xi$ must be causal. Mathematically stated, $\xi(\tau) = \theta(\tau)\xi(\tau)$, where $\theta(\tau)$ is the Heaviside step function. This requirement of causality implies that in the frequency domain, the response function $\xi(\omega)$ obeys the K.K. relation
\begin{equation}
    \text{Re}\, \xi(\omega) = -\mathcal{P}\int \frac{d\omega'}{\pi} \frac{\text{Im}\, \xi(\omega')}{\omega - \omega'}.
\end{equation}

The relevant sum rule can be developed by considering the asymptotic $\omega \to \infty$ behavior of both sides of the K.K. relation. For the left hand side, we have
\begin{equation}
    \text{Re}\, \xi(\omega)_{\omega\to\infty}  \sim \frac{1}{\omega_a + K_l''(\omega) - \omega} \sim -\frac{1}{\omega}.
\end{equation}
For the right hand side, we have
\begin{equation}
    \lim_{\omega\to\infty} -\mathcal{P}\int \frac{d\omega'}{\pi} \frac{\text{Im}\,\xi(\omega')}{\omega - \omega'} \sim -\frac{1}{\omega} \int \frac{d\omega'}{\pi} \text{Im}\,\xi(\omega').
\end{equation}
By enforcing equality of the two expressions, we obtain the sum rule
\begin{equation}
    \int d\omega\, \text{Im}\,\xi(\omega) = \pi.
\end{equation}
As a result of this sum rule, the commutation relation is now evaluated as 
\begin{equation}
    [a(t) , a^\dagger(t)] =\int \frac{d\omega}{\pi} \text{Im}\, \xi(\omega) = 1.
\end{equation}
Thus, we see that the commutation relation is preserved, so long as the coupling and loss functions are K.K. consistent with one another as to maintain causality.

\section{Steady-state mean-field behavior}

In terms of this general formalism, we will now describe how the mean-field and noise of $a$ behave in response to monochromatic coherent driving via the input field $s(t)$. Physically, this corresponds to shining light into the dispersive and nonlinear resonance via the coupling port (i.e., a waveguide), and examining the corresponding response of the resonance field. We start by decomposing the input field operator into the sum of a mean-field component, and operator valued quantum fluctuations as $s(t) = \braket{s(t)} + \delta s(t)$. Then, we assume that the input field is coherently and monochromatically driven with amplitude $s_0$ and frequency $\omega_p$, so that $\braket{s(t)} = s_0 e^{-i\omega_p t}$. Additionally, we will denote $\langle a \rangle = \alpha_0 e^{-i\omega_p t}$ as the mean-field component of $a$, which will also oscillate at the pump frequency. By substituting these assumptions into Eq.~\ref{eq:general_eom}, we find that in the steady-state, the mean-field amplitude obeys:
\begin{equation}
\begin{split}
    \frac{d}{dt}(\alpha_0 e^{-i\omega_p t}) = -i\omega_a \alpha_0 e^{-i\omega_p t} &- i\beta|\alpha_0|^2\alpha_0 e^{-i\omega_p t} \\&- \int dt' K_l(t-t') \alpha_0 e^{-i\omega_p t'} + \int dt'\,K_c(t-t')s_0 e^{-i\omega_p t'}.
\end{split}
\end{equation}
After performing Fourier manipulations, and cancelling out oscillating terms, we find the frequency domain steady-state equation
\begin{equation}
    [i(\omega_a - \omega_p) + i\beta|\alpha_0|^2 + K_l(\omega_p)]\alpha_0 = s_0 K_c(\omega_p).
\end{equation}
Thus, we see that the monochromatic drive causes the loss and coupling kernels to be sampled at the pump frequency $\omega_p$. To obtain an equation which describes the steady-state photon number $n = |\alpha_0|^2$ rather than just the field, we note that the loss function $K_l$ is complex. We will specifically write $K_l(\omega) = K_l'(\omega) + iK_l''(\omega).$ This decomposition comes with a clear interpretation, namely that $K_l'(\omega)$ encodes the frequency-dependent damping rate of the resonance $a$, while $K_l''(\omega)$ encodes the shift of the resonance frequency $\omega_a$ which accompanies the loss. In other words, the real and imaginary parts of $K_l$ respectively correspond to the dissipative and reactive dynamics induced by the frequency-dependent coupling to the environment. Using this decomposition, an equation for the steady-state resonance photon number $n$ can be written as:
\begin{equation}
    [(\omega_a + \beta n + K_l''(\omega_p) - \omega_p)^2 + K_l'(\omega_p)^2]n = |s_0|^2 |K_c(\omega_p)|^2.
    \label{eq:mean_field_steady_state}
\end{equation}
We note that via Kramers-Kronig relations between the loss and coupling kernels, $|K_c(\omega)|^2 = 2\,\text{Re}\,K_l(\omega)$. This equation of state is consistent with results known in the context of Kerr induced optical bistability \cite{drummond1980quantum}. The main difference here is that in-coupling, damping, and detuning are now frequency-dependent. In general, this means that the presence and extent of bistable behavior can now depend heavily on the frequency at which the system is driven.

\section{Steady-state quantum noise properties}

In addition to the steady-state behavior, the quantum noise properties of this driven system can also differ substantially from the behaviors seen in ordinary Kerr bistability. In this section, we will develop the theory of quantum noise in the non-Markovian driven-dissipative cavity, and develop analytical approximations which lead to the ``sharp loss'' interpretation of intensity noise reduction.

\subsection{General formalism for quantum noise}

First, we develop the general theory of quantum noise in a non-Markovian Kerr resonator. We do so through a linearization approach, which decomposes the quantum cavity and input fields into c-number mean values, plus operator valued fluctuations. For the input field, this is $s = \braket{s} + \delta s$, while for the cavity field it is $a = \braket{a} + \delta a$. By substituting this decomposition into Eq.~\ref{eq:general_eom}, and maintaining terms of only linear order in the fluctuations, we can obtain linear equations of motion for the fluctuations. In the frequency domain, the linearized operator-valued fluctuations obey the equations:
\begin{equation}
    \begin{pmatrix}
        \eta(\omega) & i\beta n \\
        -i\beta n & \eta^*(-\omega)
    \end{pmatrix} \begin{pmatrix}
        \delta a(\omega) \\ \delta a^\dagger(-\omega) 
    \end{pmatrix} = \begin{pmatrix}
        K_c(\omega_p + \omega) \delta s(\omega) \\ K_c^*(\omega_p - \omega) \delta s^\dagger (-\omega)
    \end{pmatrix}
\end{equation}
Here, we have defined $\eta(\omega) \equiv i(\omega_{ap} - \omega + 2\beta n) + K_l(\omega_p + \omega)$, where $\omega_{ap} \equiv \omega_a - \omega_p$ is the bare pump-cavity detuning. Additionally, we note that any instance of $n$ refers to the steady state mean-field dictated by Eq.~\ref{eq:mean_field_steady_state}. At a high level, this result indicates that the combination of frequency-dependent and Kerr interactions results in a frequency-dependent mixing of the two quadratures of the resonance $a$. This will result in frequency-dependent squeezing phenomena, that can result in the reduction of fluctuations below the shot noise limit. Additionally, we note that if $\delta s$ represents classical fluctuations from the mean-field steady-state solution, then the equation above can also be used to describe classical oscillations from equilibrium. A discussion of these transient behaviors is given in Section~\ref{sec:phase_diagrams}.

The matrix equation above can be inverted to solve for $\delta a(\omega)$ and $\delta a^\dagger(-\omega)$ in terms of the input fluctuations $\delta s(\omega)$ and $\delta s^\dagger(-\omega)$. We define quantities $p(\omega)$ and $q(\omega)$ such that $\delta a(\omega) = p(\omega) \delta s(\omega) + q^*(-\omega)\delta s^\dagger(-\omega)$. In terms of previously defined quantities, 
\begin{align}
    p(\omega) &= \frac{1}{M(\omega)} \eta^*(-\omega) K_c(\omega_p + \omega) \\
    q(\omega) &= \frac{i\beta n}{M(\omega)} K_c(\omega_p + \omega)
\end{align}
where we have also defined $M(\omega) \equiv \eta(\omega)\eta^*(-\omega) + (\beta n)^2$ as the determinant of the matrix acting on the fluctuations. 

In addition to the field operators $a$ and $a^\dagger$, we can also consider quadrature operators $X = a + a^\dagger$ and $Y = -i(a - a^\dagger)$. This allows us to analytically express the steady-state variance of the quadrature operators in terms of the functions $p(\omega)$ and $q(\omega)$ defined above. For example, the variance of the $X$ quadrature is given by
\begin{align}
    \braket{\delta X(t) \delta X(t)} &= \int \frac{d\omega}{2\pi}\frac{d\omega'}{2\pi}\, e^{-i(\omega + \omega')t}  \braket{\delta X(\omega) \delta X(\omega')} \\
    &= \int \frac{d\omega}{2\pi}\frac{d\omega'}{2\pi}\, e^{-i(\omega + \omega')t} \Braket{\left(\delta a(\omega) + \delta a^\dagger(-\omega)\right)\left(\delta a(\omega') + \delta a^\dagger(-\omega')\right)} \\
    &= \int\frac{d\omega}{2\pi}\frac{d\omega'}{2\pi}\, e^{-i(\omega + \omega')t} \left(p(\omega) + q(\omega)\right)\left(p^*(-\omega') + q^*(-\omega')\right) \Braket{\delta s(\omega) \delta s^\dagger(-\omega')} \\
    (\Delta X)^2 &= \int \frac{d\omega}{2\pi} |p(\omega) + q(\omega)|^2.
\end{align}
In going from the second to the third line, we have expanded $\delta a(\omega)$ and its conjugate in terms of $p(\omega)$ and $q(\omega)$, and dropped all terms which have $s$-correlators which are zero for a vacuum reservoir. Using the fact that $\Braket{\delta s(\omega) \delta s^\dagger(\omega')} = 2\pi\delta(\omega - \omega')$ for a vacuum reservoir allows us to integrate over $\omega'$ to achieve the final result. More general inputs (such as squeezed-state inputs) can be considered by considering more general input correlations.

After doing the same for $Y$, we find that the quadrature variances can be written as 
\begin{align}
    (\Delta X)^2 &= \int \frac{d\omega}{2\pi} |p(\omega) + q(\omega)|^2 \\
    (\Delta Y)^2 &= \int \frac{d\omega}{2\pi} |p(\omega) - q(\omega)|^2.
\end{align}
At this stage, no approximations have been made. As such, the evaluation of these frequency integrals result in exact results for the quadrature variances. Without loss of generaality, we choose the pump phase so that $\braket{a}$ in the steady-state is real. Then, $(\Delta X)^2$ denotes the amplitude quadrature variance, while $(\Delta Y)^2$ denotes the phase quadrature variance. The commutation relation $[a,a^\dagger]$ for the quantized resonance imposes the uncertainty relation $(\Delta X)(\Delta Y) \geq 1$ on the quadrature variances, which is a manifestation of the amplitude-phase uncertainty relationship on the cavity field. Thus, the reductions in intensity noise we consider in this work are necessarily accompanied by an increase in fluctuations in the phase quadrature. The ultimate limit of this type of nonlinear dissipation (which is formally no longer fully captured through this linearization approach) is a Fock state with perfectly defined amplitude, and a completely undefined phase. 

We can rewrite these noise expressions in forms which make them more interpretable, and also more amenable to analytical approximations. To start, we will examine the quantity $M(\omega)$, which originates from the determinant of the linearized fluctuation matrix. By using the definition of $\eta$, we can write $M(\omega) = \Omega^2(\omega) - \omega^2 + i\Gamma^2(\omega)$, where
\begin{align}
    \Omega^2(\omega) &\equiv 3(\beta n)^2 + (\omega_{ap} + \delta_-)(\omega_{ap} + \delta_+) + 2\beta n(2\omega_{ap} + \delta_+ + \delta_-) + (\delta_+ - \delta_-)\omega + \kappa_+\kappa_- \\
    \Gamma^2(\omega)&\equiv 2\beta n (\kappa_- - \kappa_+) + \kappa_-(\omega_{ap} + \delta_+ - \omega) - \kappa_+(\omega_{ap} + \delta_- + \omega).
\end{align}
We have expressed the loss function in terms of its real and imaginary parts as $K_l(\omega) = K_l'(\omega) + iK_l''(\omega)$. Additionally, we have adopted the shorthand notation $\kappa_\pm \equiv K_l'(\omega_p \pm \omega)$ (and likewise for $\delta$) to denote the loss (and detuning) evaluated around the pump frequency $\omega_p$. As a reminder, it is understood that $n$ refers to the steady-state value of $n$.

This means then that the denominator of all of these noise and commutator quantities will take the form
\begin{equation}
    \frac{1}{|M(\omega)|^2} = \frac{1}{\left[\Omega^2(\omega) - \omega^2\right]^2 + \Gamma^4(\omega)}.
\end{equation}
In this case, the $X$-quadrature variance can be expressed as
\begin{equation}
    (\Delta X)^2 = \int \frac{d\omega}{\pi} \frac{K_+'(\omega)\left[(\omega_{ap} + \beta n + \omega)^2 + K_-'^2(\omega)\right]}{\left[\Omega^2(\omega) - \omega^2\right]^2 + \Gamma^4(\omega)}.
    \label{eq:X_noise_exact}
\end{equation}
Similarly, the $Y$-quadrature variance can be expressed as
\begin{equation}
    (\Delta Y)^2 = \int \frac{d\omega}{\pi} \frac{K_+'(\omega)\left[(\omega_{ap} + 3\beta n + \omega)^2 + K_-'^2(\omega)\right]}{\left[\Omega^2(\omega) - \omega^2\right]^2 + \Gamma^4(\omega)}
    \label{eq:X_noise_exact}
\end{equation}

\subsection{Analytical approximations for quantum noise in the adiabatic limit}

Now, we focus on developing analytic expressions for the intensity quadrature variance $(\Delta X)^2$. Our approximations are based on the so-called \emph{adiabatic approximation}, which holds when the non-Markovian elements responds quickly in time compared to the resonance $a$. In this case, we find that the noise spectra are sharply peaked at frequencies $\omega_p \pm \Omega$. Here, $\Omega$ is the so-called relaxation oscillation frequency of a Markovian Kerr cavity, which is defined as $\Omega = \sqrt{\Delta^2 - (\beta n)^2}$, where $\Delta = \omega_{ap} + 2\beta n$. The other portions of the noise integral, on the other hand, are slowly varying, and thus the integral can be approximated as a pair of Lorentzian integrals peaked at $\pm\Omega$ to a high degree of accuracy. After these integrations are performed, we find an analytical expression for the $X$ quadrature variance of the cavity field:
\begin{align}
    (\Delta X)^2 &\approx \frac{1}{2\Omega \Gamma^2} \left[\kappa_+((\Delta - \beta n + \Omega)^2 + \kappa_+^2) + \kappa_- ((\Delta - \beta n - \Omega)^2 + \kappa_-^2)\right] \\
    &\approx \frac{1}{2\Omega \Gamma^2} \left[\kappa_+(\Delta - \beta n + \Omega)^2 + \kappa_- (\Delta - \beta n - \Omega)^2 \right] \label{eq:amplitude_noise_approx_1}.
\end{align}
In the above expressions, $\kappa_\pm = K_l'(\omega_p \pm \Omega)$ is the dispersive loss evaluated at the pump shifted up and down by the frequency $\Omega$. Additionally, the linewidth $\Gamma$ is defined as 
\begin{equation}
    \Gamma^2 = |\Delta (\kappa_- - \kappa_+) - \Omega (\kappa_- + \kappa_+)|
\end{equation}
In the second line of the expression for the noise, we make the approximation that the terms cubic in the loss are small compared to the detunings. This is well respected across the regimes of parameters that we consider in this work. The above expressions constitute one of our key theoretical results. As we will see, this expression for the quadrature variance depends heavily on the dispersion of the loss. In particular, we see that the dispersive loss $K_l'(\omega)$ is evaluated at two key points: $\kappa_\pm = K_l'(\omega_p \pm \Omega)$. As a reminder, $\omega_p$ is the pump frequency, and $\Omega$ is a frequency related to resonance of fluctuations, related to how the nonlinear detuning changes with $n$. Additionally, we see that the expression for $\Gamma$ (corresponding to the damping rate of the integrated peaks), depends on both the difference and the sum of the loss at the two points $\omega_p \pm \Omega$. This is the mechanism by which the noise reduction depends on the dispersive loss. Roughly speaking, the sum and difference terms present in $\Gamma$ serve as probes for the mean and slope of the dispersive loss. Down below, we will show that in cases of maximal noise reduction near the edge of a bistable branch, these interpretations become exact, yielding compact and intuitive expressions for the intensity quadrature noise. 

The expression for the noise provided by Eq.~\ref{eq:amplitude_noise_approx_1} can be cast into a more interpretable form. After algebraic manipulations, we find that it can be written as 
\begin{equation}
    (\Delta X)^2 \approx \left(1 - \frac{\beta n}{\Delta}\right)\frac{\left(\frac{\Delta}{\Omega}\right)^2 + r\left(\frac{\Delta}{\Omega}\right)}{1 + r\left(\frac{\Delta}{\Omega}\right)}.
    \label{eq:amplitude_noise_adiabatic_approx}
\end{equation}
Here, $r \equiv (\kappa_+ - \kappa_-)/(\kappa_+ + \kappa_-)$ is the ratio between the difference and sum in the loss probed at $\omega_p \pm \Omega$. Eq.~\ref{eq:amplitude_noise_adiabatic_approx} constitutes one of our key theoretical results. It describes how the amplitude noise of an externally pumped non-Markovian Kerr resonance is determined entirely by the steady state cavity photon number $n$, detuning parameters, and the dispersive loss evaluated at sideband frequencies $\pm\Omega$. 

\textbf{Important limiting cases}

Here, we discuss some important limiting cases for out analytical expression for the amplitude quadrature variance derived above.

\textbf{Same loss at both frequencies}

In this case, we have $\kappa_+ = \kappa_- \equiv \kappa$. When this is the case, we get $\Gamma^2 = 2\kappa\Omega$. We end up with
\begin{equation}
    (\Delta X)^2 = \frac{1}{2}\left[1 + \frac{\Delta - \beta n}{\Delta + \beta n} \right].
\end{equation}
The minimum possible value is $1/2$, and occurs when $\Delta = \beta n$, which occurs when $\beta n = \omega_p - \omega_a$. When these is no nonlinearity present, we have $\beta=0$ which immediately gives a quadrature variance of 1. This is what is to be expected from a linear cavity which is coherently pumped, as such a system produced a steady-state field which is in a coherent state whose two quadrature variances are equal and uncertainty limited.

\textbf{Large $n$ limit}

When $n$ becomes large, we additionally have $\Delta \to 2\beta n$, which gives
\begin{equation}
    (\Delta X)^2 \to 
    \frac{2}{3}.
\end{equation}
This limit is consistent with, for example, what has been noted in \cite{drummond1980quantum} which investigated the quantum statistical properties of optical bistability in (Markovian) Kerr cavities. In other words, some limited degree of intensity squeezing is possible in a nondisperisve Kerr cavity. On physical grounds, this effect results from different parts of the phase space distribution experiencing different rotation rates due to the intensity-dependent frequency characteristic of a cavity that contains a Kerr medium. Thus the phase space distribution of light in a pumped Kerr medium is deformed away from its symmetric coherent state form, leading to some squeezing. Note that this limit also effectively falls under the category of the losses at the two points becoming the same, because $\Omega$ grows linearly with $n$ for large $n$, and thus the dispersive loss is evaluated far away from the dispersive part. Within the context of our loss model, this means that both evaluation points are at opposite ends of the Lorentzian dip, where the dispersion is flat, and the loss values are the same.

\textbf{Sharp loss}

The expression for amplitude noise given in Eq.~\ref{eq:amplitude_noise_adiabatic_approx} allows us to make an important interpretation related to ``sharp loss'' when the system is pumped near a bistable point. In particular, the onset of a bistable point is associated with the vanishing of $\Omega$. As one approaches the bistable point, $\Omega$ thus becomes small until its eventual vanishing. When $\Omega$ becomes small relative to the scale of frequency variations in the dispersion loss $K_l'(\omega)$, this expression for the amplitude noise probes the slope of $K_l'(\omega)$. To see this we note that
\begin{equation}
    \lim_{\Omega \to 0} \frac{r}{\Omega} = \lim_{\Omega \to 0} \frac{K_l'(\omega + \Omega) - K_l'(\omega - \Omega)}{\Omega(K_l'(\omega + \Omega) + K_l'(\omega - \Omega))} = \frac{1}{K_l'(\omega_p)}\frac{dK_l'(\omega)}{d\omega}\Bigg|_{\omega=\omega_p}.
\end{equation}
Thus, as one approaches the bistable points, the noise is increasingly determined by the slope of the loss curve at the pump frequency, relative to the loss at this point. It is clear, then, that in order to maximize squeezing, one should: (1) operate with a detuning which is sufficiently large to induce bistability, (2) operate at a photon number $n$ at the lower region of the upper bistable branch, which creates a small sideband frequency $\Omega$, and (3) ensure that the ratio $\frac{1}{K_l'(\omega_p)}\frac{dK_l'(\omega)}{d\omega}\Big|_{\omega=\omega_p}$ is as large as possible, by maximizing the steepness of the dispersive loss, while minimizing all other background losses. Examples of these principles are explored further in the next section.

\subsection{Additional results for for generation of highly intensity-squeezed states}

\begin{figure}
    \centering
    \includegraphics{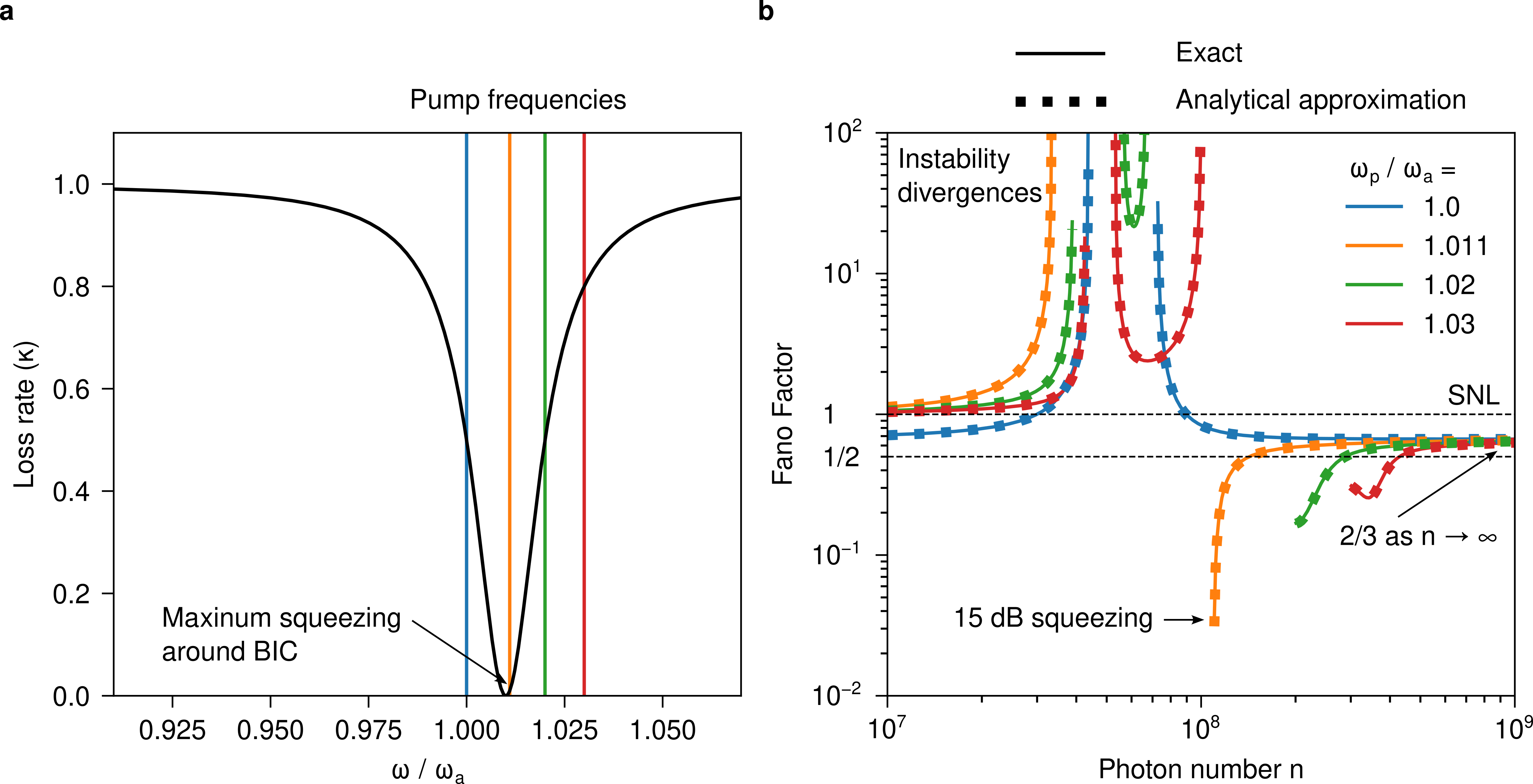}
    \caption{\textbf{Large intensity squeezing of macroscopic states.} (a) Dispersive loss curve associated with the F.W. model. Curve is marked with four pump frequencies which are represented in (b). (b) Fano factor as a function of steady state photon number for different pump frequencies $\omega_p$. A comparison between the exact numerical integral (Eq.~\ref{eq:X_noise_exact}) and the analytical approximation (Eq.~\ref{eq:amplitude_noise_adiabatic_approx}) shows very strong agreement. Parameters used are the same as those used in main text Figs. 2-3.}
    \label{fig:macroscopic_squeezing_details}
\end{figure}

\begin{figure}
    \centering
    \includegraphics{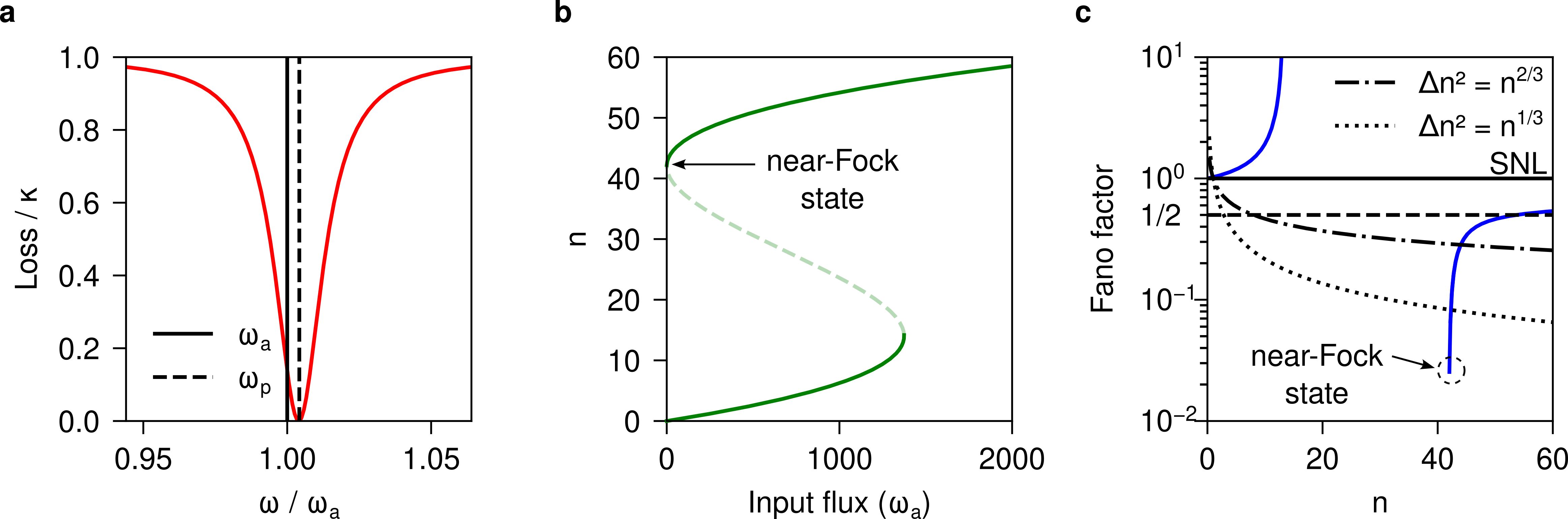}
    \caption{\textbf{Near-fock state generation with the F.W. model.} (a) Disperisve loss dip caused by interference. Values of the bare resonance $\omega_a$ and pump $\omega_p$ are shown, which are chosen to lie near the lossless mode (BIC). (b) Input-output curve which exhibits bistability for small photon numbers. (c) Amplitude noise reduction characterized by the Fano factor. Near-Fock state generation occurs around $n=42$ at the lower part of the upper bistable branch. Comparisons to other noise limits are shown. Parameters used are: $\beta = 10^{-4} \omega_a$, $\kappa = 10^{-4}\omega_a$, $\gamma = 10^{-2}\omega_a$, $\omega_d = \omega_a + 0.4 \gamma$, $\omega_p = 1.0042 \omega_a$.}
    \label{fig:FW_fock}
\end{figure}

In this section, we provide additional data for high intensity squeezing that can be achieved through non-Markovian dissipation. We first provide additional data for intensity noise reduction for the nonlinear F.W. system operated with macroscopic photon numbers (the focus of Fig. 3 of the main text). In particular, we show the amplitude noise results of main Fig. 3a, but for four different pump frequencies (Fig.~\ref{fig:macroscopic_squeezing_details}). By operating at a pump frequency just to the right of the vanishing loss point created by the BIC, it is possible to achieve a very high degree of steady state squeezing in the cavity ($> 15$ dB in this case).  

We remark on a few additional features of the amplitude noise data. First, we note that the amplitude noise exhibits divergences at lower photon numbers, due to a combination of modulational instability and ordinary bistability. Noise values are not shown in these regions, since there is no steady state for which to compute the noise. Noise reduction below the Markovian limit of $1/2$ is seen for the three pump frequencies which lie to the right of the BIC point. In these case, it is the ``sharp loss'' mechanism we have described that stabilizes these intensity-squeezed steady-states. Notably, when the pump frequency is on the left side of the BIC (where loss decreases with frequency), no such noise reduction is achieved. This pump frequency is actually below the detuning required for bistability, so there is only a single region of diverging noise, which results from modulational instability, even in the absence of bistability. Finally, all of the models approach a Fano factor of $2/3$ as $n$ increases, as described in the previous section.

\begin{figure}
    \centering
    \includegraphics{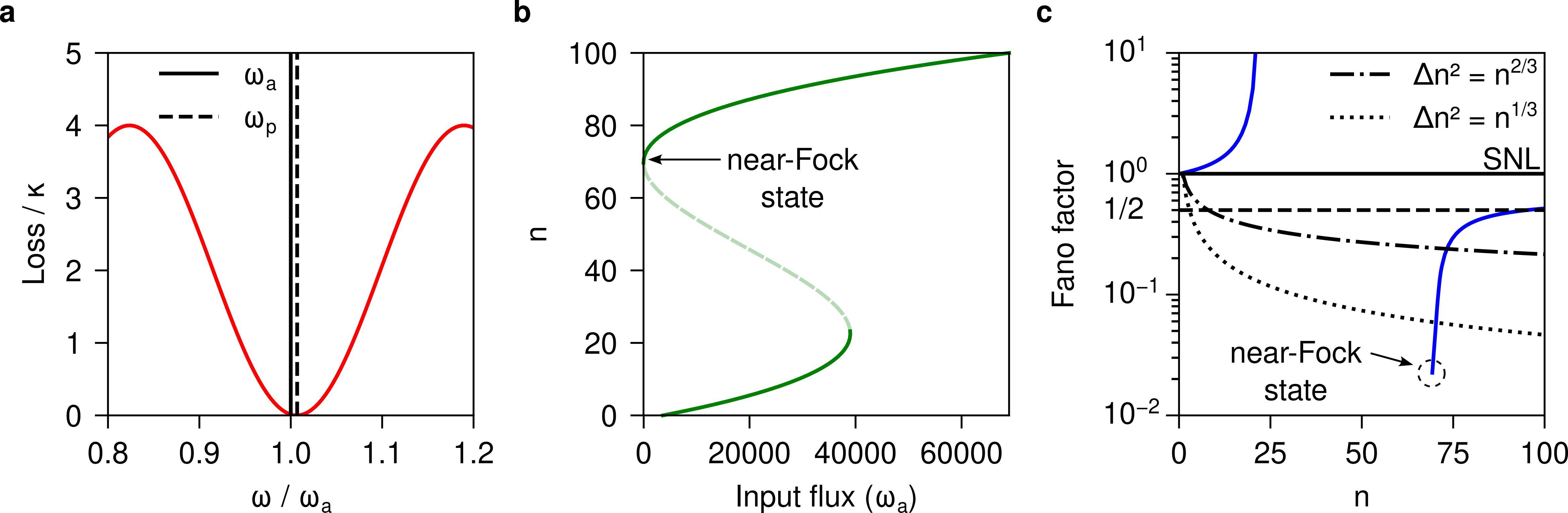}
    \caption{\textbf{Near-fock state generation with Fano mirror.} Similar setup to that shown in Fig.~\ref{fig:FW_fock}, except that the near-Fock state generation is done with a Fano mirror model. Parameters used are: $\omega_a = 1.03 \times 10^{15}$ s$^{-1}$, $\kappa = 10^{-4} \omega_a$, $L = 5$ $\mu$m, and $\beta = 10^{-4}\omega_a$, , which are chosen to represent systems with strong per-photon nonlinearity \cite{fink2018signatures}. }
    \label{fig:fano_mirror_fock}
\end{figure}

We also provide additional details on the generation of near-Fock cavity states through the coherent pumping of nonlinear non-Markovian systems. We show how this can be realized through both the nonlinear F.W. model (Fig.~\ref{fig:FW_fock}), as well as the Fano mirror model (Fig.~\ref{fig:fano_mirror_fock}). For both models, we assume a nonlinear strength $\beta = 10^{-4}\omega_a$, which is characteristic of the extreme nonlinearities found in exciton-polariton condensates \cite{fink2018signatures}. 

In both models, near-Fock states of large number (order of tens) can be created by operating around the frequency range where the losses drop due to interference. In both systems, near-Fock state generation is optimized by setting the bare cavity frequency $\omega_a$ slightly below the loss dip, while pumping at frequency $\omega_p$ slightly to the right of the loss dip. This ensures that bistable operation can occur, and that the pump frequency lies just to the right of the loss minimum, where the loss increases as a function of frequency. This enables one to exploit the ``sharp loss'' mechanism described in order to achieve the highest possible noise reduction. The lowest noise states can be created inside the cavity by operating at the very lowest part of the upper bistable branch (Figs.~\ref{fig:FW_fock}b,~\ref{fig:fano_mirror_fock}b), which can be accessed through hysteresis on the bistable input-output curve.


For the parameters chosen here, the F.W. model produces a state with $n \approx 42$ and $\Delta n \approx 1$, while the Fano mirror model produces one with $n \approx 70$ and $\Delta n \approx 1$. The non-Markovian architecture employed here gives these states a degree of amplitude squeezing which is far beyond the bounds set by other standard architectures. For starters, these states quite easily exceed the Fano factor bound $F > 1/2$ which is attainable with a conventional (Markovian) driven-dissipative cavity. Additionally, these states easily exceed the bound for the amplitude squeezing created by the displacement of squeezed vacuum ($(\Delta n)^2 \geq n^{2/3}$) \cite{bondurant1984squeezed}, and even the bound for what is possible with more elaborate Kerr squeezing architectures ($(\Delta n)^2 \geq n^{1/3}$) \cite{kitagawa1986number}. The non-Markovian driven-dissipative architecture thus presents attractive opportunities for creating high number Fock or near-Fock states which are impossible to create by more conventional means. 

The examples given here, as well as the more general physical insights provided by our analytical approximations, indicate that the generation of highly intensity-squeezed (and even near-Fock states) in nonlinear non-Markovian driven-dissipative systems is a phenomenon that can be realized quite broadly.

\section{non-Markovian loss models}
\label{sec:loss_models}

In these sections, we briefly summarize the two models of non-Markovian dissipation which are used in this work. Detailed derivations are provided in \cite{rivera2023creating}.

\subsection{Friedrich-Wintgen model}

The Friedrich-Wintgen (F.W.) model is a temporal coupled mode theory model which exhibits bound state in the continuum (BIC) formation via two resonances $a$ and $d$ coupled to a common continuum \cite{friedrich1985interfering, hsu2016bound}. We consider a quantum treatment of the F.W. model, in which one of the resonances ($a$) is additionally nonlinear. This nonlinear F.W. model thus comprises two quantized resonances, $a$ and $d$, which are coupled to a common continuum $s(t)$. The resonance $a$ is assumed to be nonlinear. In this case, the Heisenberg-Langevin equations of motion for the mode operators are:
\begin{subequations}
\begin{align}
    \dot{a} &= -i\omega_a a - i\beta (a^\dagger a)a - \kappa a - \sqrt{\kappa\gamma}d + \sqrt{2\kappa} s(t) \label{eq:nl_fw-a} \\
    \dot{d} &= -i\omega_d d  - \gamma d - \sqrt{\kappa\gamma}a + \sqrt{2\gamma} s(t) \label{eq:nl_fw-b}
\end{align}
\end{subequations}
Here, $\kappa$ and $\gamma$ are the respective decay rates of $a$ and $d$. Since $d$ obeys linear dynamics, it can be eliminated to yield an equation of motion for $a$ alone, which takes the form of Eq.~\ref{eq:general_eom}. 

For the F.W. model, the coupling and loss functions for the resonance $a$ are:
\begin{subequations}
\begin{align}
    K_c(\omega) &= \sqrt{2\kappa}  \left[1 - \frac{\gamma}{i(\omega_d - \omega) + \gamma}\right] \label{eq:fw_Kc} \\
    K_l(\omega) &= \kappa  \left[1 - \frac{\gamma}{i(\omega_d - \omega) + \gamma}\right], \label{eq:fw_Kl}
\end{align}
\end{subequations}
where $\kappa$ and $\gamma$ are the respective decay rates of $a$ and $d$, which have frequencies $\omega_{a,d}$.

\subsection{Fano mirror model}

For the Fano mirror model, we use the following coupling and loss functions to describe the dispersive dissipation of $a$:
\begin{align}
    K_l(\omega) &= 2\kappa\left(1-\frac{e^{2i\theta_2}}{r_d - e^{-i\omega T}} \right) = 2\kappa\left(1-\frac{r_d-it_d\sigma}{r_d - e^{-i\omega T}} \right) \\
    K_c(\omega) &= \sqrt{2\kappa}e^{i\theta_1}\left[1 + \frac{it_d e^{i(\theta_2-\theta_1)}}{r_d - e^{-i\omega T}} \right].
\end{align}
Here, $\kappa$ is the decay rate of $a$, $\theta_{1,2}$ are phases associated with the interference, $\sigma = \pm$ is the parity, $r_d$ and $t_d$ are the reflection and transmission coefficients of the interference, and $T$ is the round trip time of the cavity.

\section{Modulational instabilities and phase diagrams}
\label{sec:phase_diagrams}

In this section, we discuss the stability properties of non-Markovian driven-dissipative cavities. We begin with a stability analysis of the nonlinear F.W. model, and then describe how these conclusions can be more broadly generalized. 

In steady state operation, we can describe the quantum state of the field in terms of fluctuations about the steady state values. The equations of motion for the fluctuations associated with Eqs.~\ref{eq:nl_fw-a}, \ref{eq:nl_fw-b} are:
\begin{align}
    -i\omega_p \delta a + \delta \dot{a} &= (-i\omega_a - \kappa)\delta a - i\beta n(2\delta a + \delta a^\dagger) - \sqrt{\kappa\gamma} \delta d + \sqrt{2\kappa}\delta s \\
    i\omega_p \delta a^\dagger + \delta \dot{a}^\dagger &= (i\omega_a - \kappa)\delta a^\dagger + i\beta n(2\delta a^\dagger + \delta a) - \sqrt{\kappa\gamma} \delta d^\dagger + \sqrt{2\kappa}\delta s^\dagger \\
    -i\omega_p\delta d + \delta\dot{d} &= (-i\omega_d - \gamma)\delta d - \sqrt{\kappa\gamma}\delta a + \sqrt{2\gamma}\delta s \\ 
    i\omega_p\delta d^\dagger + \delta\dot{d}^\dagger &= (i\omega_d - \gamma)\delta d^\dagger - \sqrt{\kappa\gamma}\delta a^\dagger + \sqrt{2\gamma}\delta s^\dagger.
\end{align}
The two resonances can also be described in terms of quadrature operators $X_{a} = a + a^\dagger$, $Y_a = -i(a - a^\dagger)$ (and likewise for $d$). Then fluctuations in the quadratures from their mean values (for example, $X_a = \braket{X_a} + \delta X_a$). The linearized equation of motion for these quadratures in the case of the nonlinear F.W. model are given as:
\begin{equation}
    \frac{d}{dt}\begin{pmatrix}
        \delta X_a \\ \delta Y_a \\ \delta X_d \\ \delta Y_d
    \end{pmatrix} = \begin{pmatrix}
        -\kappa & \omega_{ap} + \beta n & -\sqrt{\kappa\gamma} & 0 \\
        -\omega_{ap} - 3\beta n & -\kappa & 0 & -\sqrt{\kappa\gamma} \\
        -\sqrt{\kappa\gamma} & 0 & -\gamma & \omega_{dp} \\
        0 & -\sqrt{\kappa\gamma} & -\omega_{dp} & -\gamma
    \end{pmatrix}\begin{pmatrix}
        \delta X_a \\ \delta Y_a \\ \delta X_d \\ \delta Y_d
    \end{pmatrix} + \begin{pmatrix}
        F_X^a \\ F_Y^a \\ F_X^d \\ F_Y^d
    \end{pmatrix}
    \label{eq:nonlinear_FW_quadrature_fluctuations}
\end{equation}
The source terms on the right hand side are operator-valued Langevin forces, which are defined as:
\begin{align}
    F_X^a &= \sqrt{2\kappa}\left(\delta s + \delta s^\dagger\right) \\
    F_Y^a &= -i\sqrt{2\kappa}\left(\delta s - \delta s^\dagger\right) \\
    F_X^d &= \sqrt{2\gamma}\left(\delta s + \delta s^\dagger\right) \\
    F_Y^d &= -i\sqrt{2\gamma}\left(\delta s - \delta s^\dagger\right).
\end{align}
By finding the eigenvalues of the matrix in the above equations of motion, we can determine the nature of the stability. Although Eq.~\ref{eq:nonlinear_FW_quadrature_fluctuations} has been developed in the full quantum picture, it can also be reduced to a classical limit by taking expectation values of both sides. In this case, the Langevin forces (which have zero expectation value) will vanish, leaving only a a classical linear matrix equation with no sources. Such an equation governs the linearized dynamics of the quadrature variables away from the mean-field steady-state behavior.

\begin{figure}
    \centering
    \includegraphics{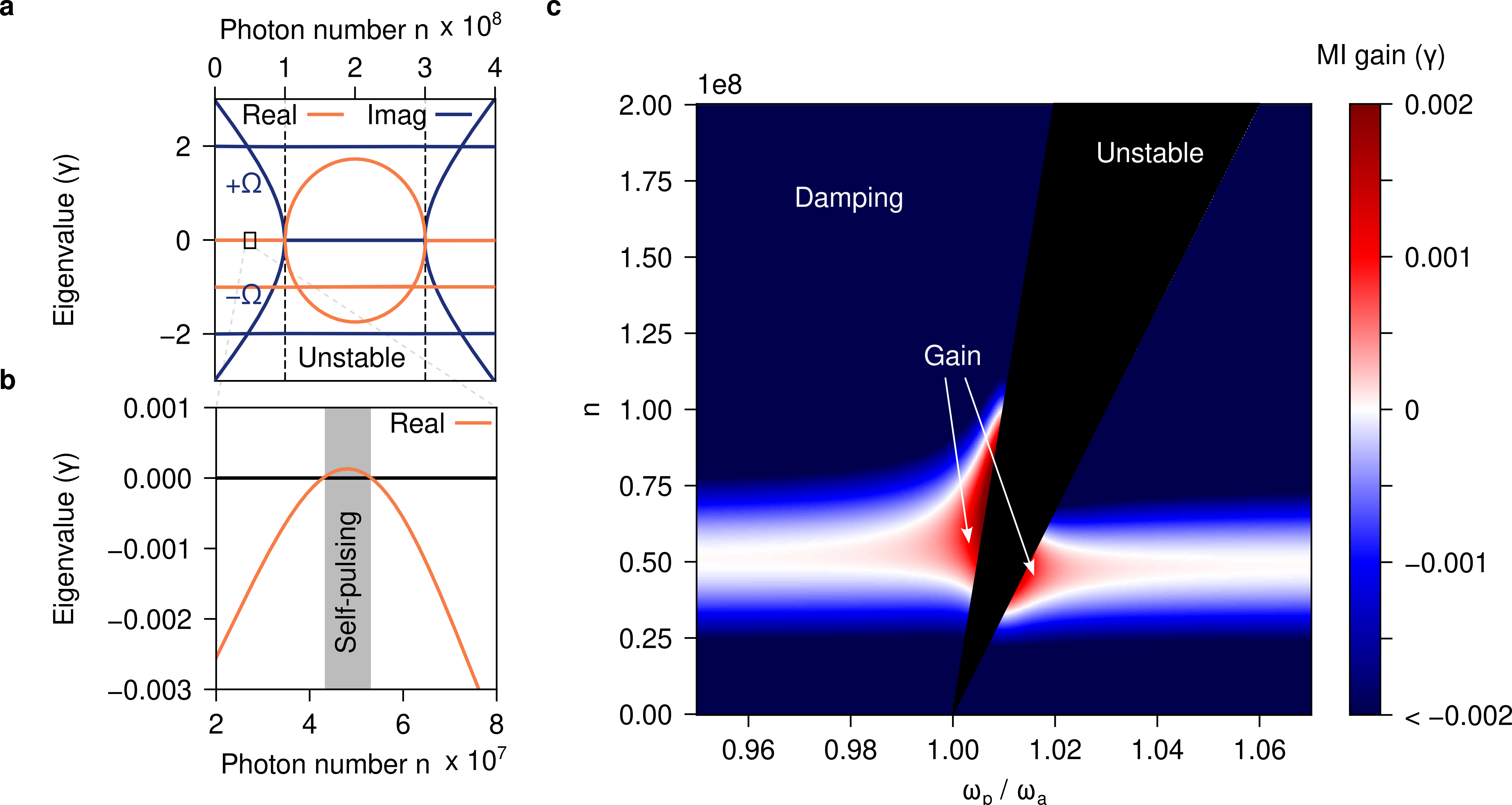}
    \caption{\text{Stability eigenvalues in the nonlinear F.W. model.} (a) Complex eigenvalues of the nonlinear F.W. model as a function of steady state photon number $n$. (b) Zoom in of the small region of (a) over which one of the eigenvalues acquires a small positive real part, leading to modulational instability. (c) Phase diagram of the nonlinear F.W. model which shows the amount of MI gain as a function of pump frequency $\omega_p$ and photon number $n$. Gain is seen to be highest for regions of phase space in proximity to the conventional unstable region.}
    \label{fig:MI_gain}
\end{figure}

In Fig.~\ref{fig:MI_gain}, we provide more details about the eigenvalues of the nonlinear F.W. model, obtained by diagonalizing the matrix in Eq.~\ref{eq:nonlinear_FW_quadrature_fluctuations}. Each of the eigenvalues may have an imaginary part $\Omega$, and a real part $\Gamma$, which dictate the decaying (or growing) oscillations which occur when the system is perturbed from equilibrium. In particular, $\Gamma > 0$ is taken to refer to damped oscillations, while $\Gamma < 0$ is taken to refer to growing oscillations. Fig.~\ref{fig:MI_gain}a shows the real and imaginary parts of the eigenvalues as a function of steady state photon number $n$. The region in the middle marked as ``unstable'' has an eigenvalue with positive real part, while $\Omega = 0$. This region is unstable due to conventional cavity bistability, and trajectories in this region will be pushed away from the region. It can also be seen that one of the eigenvalue pairs have imaginary parts $\pm\Omega$ which approach zero near these boundaries. 

The region of photon numbers which give rise to modulational instability is highlighted in Fig.~\ref{fig:MI_gain}b. This region hosts a small band of photon numbers over which $\Gamma < 0$, corresponding to growth of fluctuations. Fig.~\ref{fig:MI_gain}c shows the MI gain $-\Gamma$ in the context of the whole phase space of $n$ and $\omega_p$. We see that damping occurs in the majority of the phase diagram (blue), while modulational instability gain occurs over two bands which emerge from the conventional unstable region. Moreover, the MI gain is highest in the vicinity of the conventional unstable region, and tapers off toward zero with larger detunings from $\omega_p$ in either direction.

\begin{figure}
    \centering
    \includegraphics{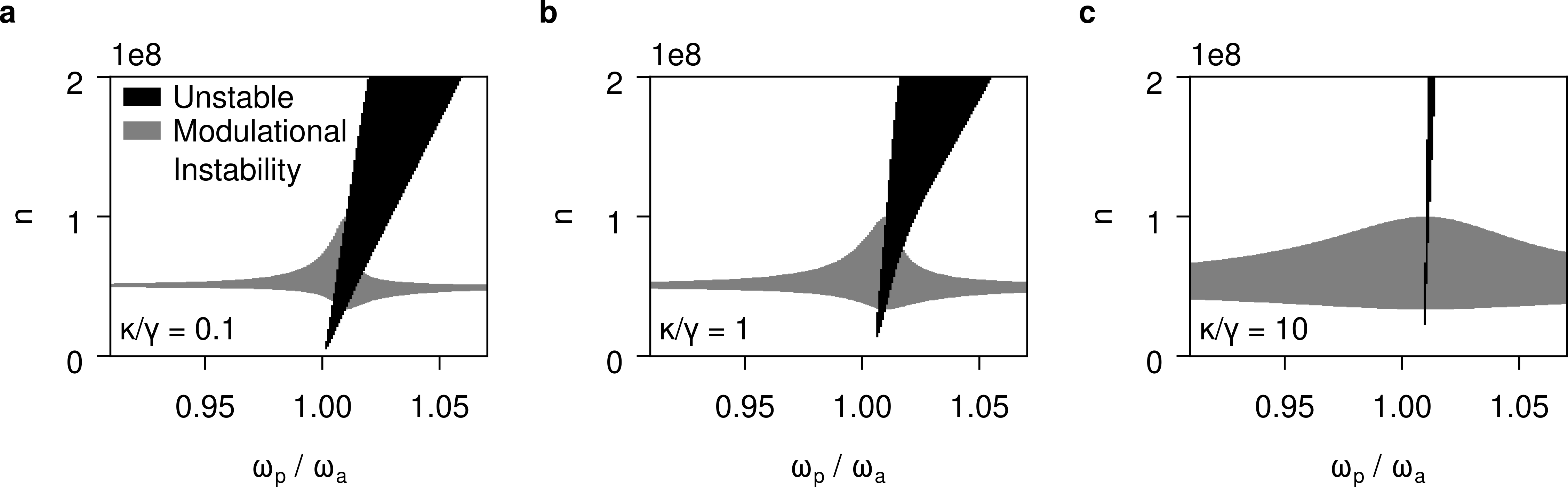}
    \caption{\textbf{F.W. model phase diagrams for different adiabatic ratios.} Phase diagrams of the driven nonlinear F.W. model are shown. Parameters used are: $\beta = 10^{-10}\omega_a$, $\gamma = 10^{-2}\omega_a$, $\omega_d = \omega_a + \gamma$. Values of $\kappa$ are varied to produce the adiabatic ratios $\kappa/\gamma = \{0.1, 1, 10\}$ marked in the panels.}
    \label{fig:adiabatic_ratio}
\end{figure}

In Fig.~\ref{fig:adiabatic_ratio}, we show how the phase diagram of the nonlinear F.W. model depends on the leakage rates $\kappa$ and $\gamma$. For the three panels shown, $\gamma$ is fixed, while $\kappa$ is varied. All three sets of parameters have regions of instability due to the traditional bistability, as well as the self-pulsing induced by modulational instability (MI).

\textbf{More general models}

In the adiabatic limit, we find that the the real part of the larger eigenvalue is well-approximated by
\begin{equation}
    \text{Re}\,\lambda = -\frac{\kappa_+ + \kappa_-}{2} - \frac{\Delta}{\Omega}\frac{\kappa_+ - \kappa_-}{2}.
\end{equation}
We thus see that this can vanish, and thus change sign, when 
\begin{equation}
    1 + \frac{\Delta}{\Omega}\frac{\kappa_+ - \kappa_-}{\kappa_- + \kappa_+} = 0.
\end{equation}
This is exactly the condition see in the denominator of the noise expressions in the adiabatic limit (see Eq.~\ref{eq:amplitude_noise_adiabatic_approx}). Thus, this expression can be used to determine the boundaries of modulational instability, even in more general cases, so long as the adiabatic approximation is respected. This is how we have determined the phase diagram in the Fano mirror model (Fig. 4 of the main text). We find the MI boundaries predicted by this method to be in strong agreement with those which are exactly determined by observing where the amplitude and phase noise diverge using the exact expression for the noise (Eq.~\ref{eq:X_noise_exact}), as evaluated by numerical integration.

\bibliographystyle{unsrt}
\bibliography{supp.bib}